\documentclass[a4paper, 11pt, eqs]{article}
\usepackage{latexsym}

\def\cleq{\setcounter{equation}{0}}

\makeatletter
\newcommand\xleftrightarrow[2][]{%
  \ext@arrow 9999{\longleftrightarrowfill@}{#1}{#2}}
\newcommand\longleftrightarrowfill@{%
  \arrowfill@\leftarrow\relbar\rightarrow}
\makeatother

\title{
T-duality as coordinates permutation in double space for weakly curved background
\thanks{Work supported in part by
the Serbian Ministry of Education and Science, under contract No. 171031.}}
\author{B. Sazdovi\'c
\thanks{e-mail: sazdovic@ipb.ac.rs}\\
{\it Institute of Physics,}\\
{\it University of Belgrade,}\\
{\it 11001 Belgrade, P.O.Box 57, Serbia}}

\begin{document}
\maketitle
\begin{abstract}
In the paper \cite{SB} we showed that in double space, where all initial coordinates $x^\mu$ are doubled  $x^\mu \to y_\mu$, the T-duality transformations can be performed by exchanging places
of some coordinates $x^a$ and corresponding dual coordinates $y_a$.
Here we generalize this result to the case of weakly curved background where in addition to the extended coordinate we will also transform extended
argument of background fields with the same operator  $\hat {\cal T}^a$.
So, in the weakly curved background T-duality leads to the physically equivalent theory and complete set of T-duality transformations form
the same group as in the flat background.
Therefore, the double space represent  all T-dual theories in unified manner.

\end{abstract}

\section{Introduction}

The T-duality  is one of the stringy properties, because it has no analogy in particle physics. Its distinguishing features are unification of equations of motion with Bianchi identity.
The standard way to construct T-dual theory is Buscher's proscription \cite{B,RV,GPR,AABL}.
In order to apply such approach it is necessary that background has some continuous isometries which leaves the action invariant. Then, in some adopted coordinates, the
background does not depend on these coordinates.
For the backgrounds which depend on the coordinates such approach is not applicable.

The simplest coordinate depending background is the weakly curved background. There the metric $G_{\mu \nu}$ is constant and the Kalb-Ramond field $B_{\mu \nu}$ is linear in coordinates with infinitesimal
coefficient. In the paper \cite{DS1} the new procedure for T-duality, adopted for the case of the  weakly curved background, has been introduced.
This  approach   generalize Buscher's one and  makes it possible to carry out T-duality along coordinates on which the Kalb-Ramond field depends.
In that article  T-duality transformations has been performed simultaneously along all coordinates $x^\mu : T^{full}=T^0\circ T^1\circ \dots \circ  T^{D-1} \,\, ,(\mu=0,1,\cdots , D)$, while
in the article \cite{DNS2} it has been performed along any subset of coordinates $x^a : T^a=T^0\circ T^1\circ \dots \circ  T^{d-1}  \,\, ,(a=0,1,\cdots , d-1)$.
The first case  connects  the beginning and the end of the T-duality chain,
\begin{eqnarray}\label{eq:chain2}
&&\Pi_{\pm \mu \nu},\,x^{\mu}
\mathrel{{\mathop{\mathop{\rightleftharpoons}^{\mathrm{T^1}}_{\mathrm{T_1}}}}}
\Pi_{1 \pm \mu \nu},\,x_1^{\mu}
\mathrel{{\mathop{\mathop{\rightleftharpoons}^{\mathrm{T^2}}_{\mathrm{T_2}}}}}
\Pi_{2 \pm \mu \nu},\,x_2^{\mu}
\dots
\mathrel{{\mathop{\mathop{\rightleftharpoons}^{\mathrm{T^i}}_{\mathrm{T_i}}}}}
\Pi_{i \pm \mu \nu},\,x_i^{\mu}
\dots
\mathrel{{\mathop{\mathop{\rightleftharpoons}^{\mathrm{T^D}}_{\mathrm{T_D}}}}}
\Pi_{D \pm \mu \nu},\,x_D^{\mu}  \, ,
\end{eqnarray}
while the second one connects  the beginning with the arbitrary node.
Here $\Pi_{i \pm \mu \nu}$ and $x^\mu_{i},\,(i=1,2,\cdots,D)$ are background fields and the coordinates of the corresponding configurations and we will also use notation $\Pi_{D \pm \mu \nu}={}^\star \Pi_{\pm \mu \nu}$ and $x_D^{\mu}=y_\mu$.
The nontrivial extension of T-duality transformations in this approach, compared with the flat space case, is a source of closed string non-commutativity
\cite{Lust,ALLP,DNS}.

The  T-duality in the extended space has been investigated in Refs.\cite{Duff}-\cite{Hull2}. In Ref.\cite{Duff} all coordinates are doubled and T-duality relation between the beginning and the end of the chain has been established for the
flat space. In Ref.\cite{Hull} only coordinates along which T-duality is performed are doubled and background fields do not depend on them.
The relation with our approach has been discussed in Ref.\cite{SB}.

In paper \cite{SB} the extended space  with coordinates $Z^M= (x^\mu, y_\mu)$, which contains all the coordinates of the initial and T-dual spaces, has been introduced.
It was shown that in such double space T-duality has a simple interpretation.
Arbitrary T-duality $ {\cal T}^a=T^a \circ T_a $,
along some initial coordinates
$x^a : T^a=T^0\circ T^1\circ \dots \circ  T^{d-1}$, and along corresponding T-dual  ones $y_a : T_a=T_0\circ T_1\circ \dots \circ  T_{d-1} $,
can be realized by exchanging their places, $x^a \leftrightarrow  y_a$.
It has been proven for constant background  fields, the metric $G_{\mu \nu}$ and the Kalb-Ramond field $B_{\mu \nu}$,
when Buscher's approach can be applied. This interpretation shows that T-duality leads to the equivalent theory, because replacement of coordinates
does not change the physics.

In the present article, following idea of Ref.\cite{SB}, we are going to offer similar interpretation of T-duality in the weakly curved background.
The main difference, comparing with the flat space case, is that in the weakly curved background the background fields depend on the coordinates. So, together with changing the coordinates, we should change the arguments of
the background fields, also.

Let us stress that we doubled all the coordinates. We rewrite T-duality  transformations connected
beginning and end of the chain (\ref{eq:chain2}) in the double space. We obtain the fundamental expression, where the generalized metric
depend on both initial and T-dual coordinates. We will show that this expression is enough to find background fields from all nodes of the chain (\ref{eq:chain2}) and T-duality
transformations between arbitrary nodes.
In such a way, as well as in the flat background, we unify all T-dual theories of the chain (\ref{eq:chain2}).

In Sec.6 we will illustrate our approach by the example  of three torus.

\cleq

\section{T-duality in the weakly curved background}

The propagation of the closed bosonic string  in D-dimensional space-time is described by the action \cite{S}
\begin{equation}\label{eq:action0}
S[x] = \kappa \int_{\Sigma} d^2\xi\sqrt{-g}
\Big[\frac{1}{2}{g}^{\alpha\beta}G_{\mu\nu}[x]
+\frac{\epsilon^{\alpha\beta}}{\sqrt{-g}}B_{\mu\nu}[x]\Big]
\partial_{\alpha}x^{\mu}\partial_{\beta}x^{\nu},
\quad (\varepsilon^{01}=-1) \, .
\end{equation}
Here $x^{\mu}(\xi),\ \mu=0,1,...,D-1$ are the coordinates of the string moving in the background,
defined by the space-time metric $G_{\mu\nu}$ and the Kalb-Ramond field $B_{\mu\nu}$.
The intrinsic world-sheet metric we denote by  $g_{\alpha\beta}$. The integration goes over two-dimensional world-sheet $\Sigma$
with coordinates $\xi^\alpha$ ($\xi^{0}=\tau,\ \xi^{1}=\sigma$).

The space-time  equations of motion, in the lowest order in slope parameter $\alpha^\prime$,
for the constant dilaton field $\Phi=const$  have the form
\begin{equation}\label{eq:ste}
R_{\mu \nu} - \frac{1}{4} B_{\mu \rho \sigma}
B_{\nu}^{\ \rho \sigma}=0\, ,  \qquad
D_\rho B^{\rho}_{\ \mu \nu} = 0   \, .
\end{equation}
Here $B_{\mu\nu\rho}=\partial_\mu B_{\nu\rho} +\partial_\nu B_{\rho\mu}+\partial_\rho B_{\mu\nu}$ is the field strength of the field $B_{\mu \nu}$, and
$R_{\mu \nu}$ and $D_\mu$ are Ricci tensor and covariant derivative with respect to space-time metric. The equations of motion are consequence of the world-sheet conformal invariance on the quantum level.

We will consider the simplest coordinate dependent solutions
of (\ref{eq:ste}), the so-called weakly curved background, defined as
\begin{equation}
G_{\mu\nu}=const,
\quad
B_{\mu\nu}[x]=b_{\mu\nu}+ \frac{1}{3}B_{\mu\nu\rho}x^\rho \equiv  b_{\mu\nu}+ h_{\mu\nu}(x) \,  .
\end{equation}
This background satisfies the space-time equations of motion, if the constant
$B_{\mu\nu\rho}$ is taken to be infinitesimally small (for more details see \cite{ALLP}). Then all the calculations can be performed  in the first order
in $B_{\mu\nu\rho}$, when the Ricchi tensor can be neglected as the infinitesimal of the second order.

In the conformal gauge $g_{\alpha\beta}=e^{2F}\eta_{\alpha\beta}$, and  light-cone coordinates
$\xi^{\pm}=\frac{1}{2}(\tau\pm\sigma)$, $\partial_{\pm}= \partial_{\tau}\pm\partial_{\sigma}$,
the action (\ref{eq:action0}) obtains the form
\begin{equation}\label{eq:action1}
S[x] = \kappa \int_{\Sigma} d^2\xi\
\partial_{+}x^{\mu}
\Pi_{+\mu\nu}[x]
\partial_{-}x^{\nu},
\end{equation}
where we introduced the useful combination of the background fields
\begin{eqnarray}
\Pi_{\pm\mu\nu}[x]=
B_{\mu\nu}[x]\pm\frac{1}{2}G_{\mu\nu}.
\end{eqnarray}

We will assume that background has topology of $D$-dimensional torus, $T^D$. In the Section 6,  we will present example of the
$3$-dimensional torus, $T^3$.


\subsection{Sigma-model approach to T-duality in the weakly curved background}

T-dualization along all the coordinates in the weakly curved background, has been obtained in Ref.\cite{DS1}.
Let us stress that the coordinates of which the background fields do not depend on, presents Killing directions. The usual Buscher
procedure can be applied along these directions. Here, following Ref.\cite{DS1}, we will consider the more general approach  when Kalb-Ramond field depend on some coordinates. Then, in the case of weakly curved background variation of the action with respect to the argument of Kalb-Ramond field is total divergence. For topologically trivial mapping of the world-sheet into space-time it vanishes. It means that classically, directions which appear in the argument of Kalb-Ramond field are also Killing directions, but now the usual Buscher procedure can not be applied. The explanation is that the  argument of Kalb-Ramond field depend only on the coordinates and not on its derivatives. In order to generalize  Buscher procedure we must find the gauge invariant coordinate. It is nontrivial step and have been realized in Refs.\cite{DS1,DNS2}.

In Ref.\cite{DS1} we obtaind the T-dual action
\begin{equation}\label{eq:dualna}
S[y]=
\kappa
\int d^{2}\xi\
\partial_{+}y_\mu
\,^\star \Pi_{+}^{\mu\nu}\big(\Delta V(y)\big)\,
\partial_{-}y_\nu
=\,
\frac{\kappa^{2}}{2}
\int d^{2}\xi\
\partial_{+}y_\mu
\theta_{-}^{\mu\nu}\big (\Delta V(y) \big)
\partial_{-}y_\nu,
\end{equation}
where
\begin{eqnarray}
{\theta}^{\mu\nu}_{\pm} (\Delta V) &\equiv&
-\frac{2}{\kappa}
(G^{-1}_{E}\Pi_{\pm}G^{-1})^{\mu\nu}=
{\theta}^{\mu\nu}\mp \frac{1}{\kappa}(G_{E}^{-1})^{\mu\nu} \,  .
\end{eqnarray}
Its symmetric and antisymmetric parts are the inverse of the effective metric $G^E_{\mu\nu}$ and the non-commutativity parameter $\theta^{\mu\nu}$
\begin{eqnarray}
G^E_{\mu\nu}  (\Delta V) & \equiv  G_{\mu\nu}-4(BG^{-1}B)_{\mu\nu}, \qquad
\theta^{\mu\nu}  (\Delta V) & \equiv  -\frac{2}{\kappa} (G^{-1}_{E}BG^{-1})^{\mu\nu}.
\end{eqnarray}
They depend on the expression
\begin{equation}\label{eq:vargdef}
\Delta V^\mu(y)
=-\kappa\theta_{0}^{\mu\nu}\Delta y_\nu+(g^{-1})^{\mu\nu}\Delta{\tilde{y}}_\nu,
\end{equation}
where
$\Delta y_\mu=y_\mu(\xi)-y_\mu(\xi_{0})$ and
\begin{equation}\label{eq:dualnavar}
\Delta{\tilde{y}}_\mu=\int(d\tau y^\prime_\mu+d\sigma \dot{y}_\mu).
\end{equation}
We also introduced  flat space effective metric and non-commutativity parameter
$g_{\mu\nu}=(G -4b G^{-1}b)_{\mu\nu}$ and ${\theta}^{\mu\nu}_{0} =-\frac{2}{\kappa} (g^{-1}bG^{-1})^{\mu\nu}$ as well as their combinations
${\theta}^{\mu\nu}_{0 \pm} =  {\theta}^{\mu\nu}_{0} \mp -\frac{1}{\kappa} (g^{-1})^{\mu\nu}$.

Consequently, both T-dual background fields  in the case of weakly curved background depend on the coordinates  $\Delta V (y)$ and
have a form
\begin{equation}\label{eq:tdbf}
^\star G^{\mu\nu}\big[\Delta V(y)\big]=
(G_{E}^{-1})^{\mu\nu}\big[\Delta V(y)\big],
\quad
^\star B^{\mu\nu}\big[\Delta V(y)\big]=
\frac{\kappa}{2}
{\theta}^{\mu\nu}\big[\Delta  V(y)\big]  \, .
\end{equation}
Note that the dual effective metric is inverse of the initial metric and hence it is coordinate independent
\begin{eqnarray}\label{eq:cdi1}
{}^\star G_E^{\mu\nu} \equiv {}^\star G^{\mu\nu}-4({}^\star B {}^\star G^{-1} {}^\star B)^{\mu\nu}= (G^{-1})^{\mu\nu}  \,  ,
\end{eqnarray}
while the following combination depend on the coordinates
\begin{equation}\label{eq:cdi2}
({}^\star B {}^\star G^{-1})^\mu{}_\nu = - (G^{-1} B)^\mu{}_\nu  \,  , \qquad
({}^\star G^{-1} {}^\star B)_\mu{}^\nu = -(B G^{-1})_\mu{}^\nu  \,  .
\end{equation}

The fact that we work with the weakly curved background ensures that T-dual theory is solution of the space-time equations
(\ref{eq:ste}). Because both dual metric ${}^\star G^{\mu\nu}$ and dual Kalb-Ramond field ${}^\star B^{\mu \nu}$ are linear in coordinates with infinitesimal coefficients, then
dual Christoffel  ${}^\star \Gamma_\mu^{\nu \rho}$ and dual field strength  ${}^\star B^{\mu \nu \rho}$ are constant and infinitesimal.
So, both dual space-time equations, for the metric and for the Kalb-Ramond field,  are infinitesimal of the second order which we will neglect.


\subsection{T-duality transformations in the weakly curved background}

As well as in Ref.\cite{SB} we will start with the T-duality transformations between all initial coordinates $x^\mu$ and all T-dual coordinates $y_\mu$.
For the closed string propagating  in the weakly curved background they have been derived in ref.\cite{DS1}
\begin{eqnarray}\label{eq:xtdual}
\partial_{\pm}x^\mu \cong
-\kappa\theta^{\mu\nu}_{\pm}[\Delta V] \Big{[}
\partial_{\pm} y_\nu
\pm 2\beta^{\mp}_{\nu} [V] \Big{]} \,  , \nonumber \\
\partial_{\pm}y_\mu\cong
-2\Pi_{\mp\mu\nu}[x]\partial_{\pm}x^\nu\mp
2\beta^{\mp}_{\mu}[x].
\end{eqnarray}
Here $V^\mu$ is defined in (\ref{eq:vargdef}) and the functions $\beta^\pm_\mu$ have a form
\begin{equation}\label{eq:beta}
\beta^\pm_\mu[x]=\frac{1}{2}(\beta^0_\mu\pm\beta^1_\mu)=
\mp\frac{1}{2} h_{\mu\nu}[x]\partial_{\mp} x^\nu,\,
\beta^0_\mu[x]=h_{\mu\nu}[x]x'^\nu,\,
\beta^1_\mu[x]=-h_{\mu\nu}[x]\dot x^\nu\, .
\end{equation}

If $B_{\mu \nu}(x)$ does not depend on some coordinate $x^{\mu_1}$, then the corresponding $\beta_{\mu_1}$ functions are equal to zero,
$\beta^\pm_{\mu_1}= \beta^0_{\mu_1}=\beta^1_{\mu_1}=0$. Because in that case the standard Buscher approach can be applied, from now on we
will suppose that $B_{\mu \nu}(x)$ depend on all coordinates.

The transformations  (\ref{eq:xtdual}) are inverse to one another.
Using the fact that
\begin{equation}
\theta^{\mu\nu}_{\pm}(x) = \theta^{\mu\nu}_{0\pm} -2\kappa
[\theta_{0\pm}h(x)\theta_{0\pm}]^{\mu\nu},
\end{equation}
we can reexpress these T-duality transformations as
\begin{eqnarray}\label{eq:xtdualdi}
\partial_{\pm}x^\mu \cong
-\kappa  \, {}^{\diamond} \theta^{\mu\nu}_{\pm}[\Delta V] \partial_{\pm} y_\nu
 \,  , \nonumber \\
\partial_{\pm}y_\mu\cong
-2 \, {}^{\diamond} \Pi_{\mp\mu\nu}[ x]\partial_{\pm}x^\nu  \,  .
\end{eqnarray}
Here with diamond we denoted redefined background fields, where infinitesimally small parts are rescaled
\begin{eqnarray}\label{eq:risp}
{}^{\diamond} B_{\mu \nu}(x) \equiv  b_{\mu \nu} + \frac{3}{2} h_{\mu \nu}(x) \, , \nonumber \\
{}^{\diamond} G_{\mu \nu}^E(x) \equiv g_{\mu \nu} + \frac{3}{2}\Delta G_{\mu \nu}^E(x) \, ,
\end{eqnarray}
and $\Delta G_{\mu \nu}^E(x) = G_{\mu \nu}^E(x) -g_{\mu \nu} = -4 [bh(x)+h(x)b]_{\mu \nu}$.
Let us explain the origin of the coefficient $\frac{3}{2}$. The T-duality relations (\ref{eq:xtdual}) have been obtained varying
the gauge fixed action  of ref.\cite{DS1} with respect to $v_\pm^\mu$. From the quadratic terms in $v_\pm^\mu$ we obtain the coefficient $2$ and from that of third degree
in $v_\pm^\mu$ we obtain the coefficient $3$.  Note that equality of some redefined background fields is equivalent to the
equality of corresponding initial background fields
\begin{equation}\label{eq:ebf}
{}^{\diamond} B_1(x) = {}^{\diamond} B_2(x) \,   \Longleftrightarrow  \,    B_1(x) = B_2(x)  \,  ,
\end{equation}
because both finite and infinitesimal parts are equal. Similarly, all relations between background fields as  (\ref{eq:cdi1}) and (\ref{eq:cdi2})
also valid with diamond.

Finally, we can rewrite above T-duality transformations in a form
\begin{eqnarray}\label{eq:tdualc}
& \pm \partial_{\pm}y_\mu \cong {}^{\diamond} G^E_{\mu \nu} (V) \partial_{\pm}x^\nu
-2 [{}^{\diamond} B(V)G^{-1}]_\mu{}^\nu \partial_{\pm}y_\nu \, ,  \nonumber \\
& \pm \partial_{\pm}x^\mu \cong 2 [G^{-1} {}^{\diamond} B(x)]^\mu {}_\nu \partial_{\pm}x^\nu
+(G^{-1})^{\mu \nu} \partial_{\pm}y_\nu \, ,
\end{eqnarray}
where the terms with world-sheet antisymmetric tensor
$\varepsilon_\alpha{}^\beta$ ($\varepsilon_\pm{}^\pm = \pm 1$) are  on the left hand side.
In the  double space, which contains all initial and T-dual coordinates
\begin{equation}\label{eq:escoor}
Z^M=\left (
\begin{array}{c}
 x^\mu  \\
y_\mu
\end{array}\right )\, ,
\end{equation}
these T-duality relations obtain the simple form
\begin{equation}\label{eq:tdual}
\partial_{\pm} Z^M \cong \pm \, \Omega^{MN} {\cal{H}}_{NK} (x,V) \,\partial_{\pm}Z^K \, .
\end{equation}
Here we introduced
\begin{equation}
\Omega^{MN}= \left (
\begin{array}{cc}
0 &  1 \\
1  & 0
\end{array}\right )\, ,
\end{equation}
and the coordinate dependent generalized metric
\begin{equation}\label{eq:gm}
{\cal{H}}_{MN} (x,V) = \left (
\begin{array}{cc}
{}^{\diamond} G^E_{\mu \nu}(V) &  -2 \, {}^{\diamond} B_{\mu\rho}(V) (G^{-1})^{\rho \nu}  \\
2 (G^{-1})^{\mu \rho} \, {}^{\diamond} B_{\rho \nu}(x) & (G^{-1})^{\mu \nu}
\end{array}\right )\, .
\end{equation}
Using (\ref{eq:cdi1}) and (\ref{eq:cdi2}) we can rewrite the generalized metric in terms of T-dual background fields
\begin{equation}\label{eq:gmd}
{\cal{H}}_{MN} (x,V) = \left (
\begin{array}{cc}
({}^{\diamond \star} G^{-1})_{\mu \nu}(V)                        &      2 ({}^{\diamond \star} G^{-1})^{\mu \rho} (V) \, {}^{\diamond \star} B_{\rho \nu} (V)         \\
-2 \, {}^{\diamond \star} B_{\mu\rho} (x) ({}^{\diamond \star} G^{-1})^{\rho \nu} (x)    &                ({}^{\diamond \star}  G_E)^{\mu \nu} (x)
\end{array}\right )\, .
\end{equation}

In Double field theory \cite{HZ}-\cite{BT} it is usual to call $\Omega^{MN}$ the $O(D,D)$ invariant metric and denote with $\eta^{MN}$.

The argument of the generalized metric was not written manifestly in double form. We can rewrite generalized metric as
\begin{equation}\label{eq:gmdf}
 {\cal{H}}_{NK} (x,V) = {\cal{H}}^0_{NK} + Z^M_{arg} H_{MNK} \equiv  {\cal{H}}_{NK} (Z_{arg})  \, ,
\end{equation}
where the zeroth order generalized metric
\begin{equation}\label{eq:h0}
{\cal{H}}^{(0)}_{MN} = \left (
\begin{array}{cc}
 g_{\mu \nu}  &  -2  (b G^{-1})_\mu{}^\nu   \\
2 (G^{-1}b)^\mu {}_\nu  & (G^{-1})^{\mu \nu}
\end{array}\right )\, ,
\end{equation}
and infinitesimal coefficient
\begin{equation}\label{eq:fsgm}
{\cal{H}}_{MNK} = \left (
\begin{array}{cc}
-2[b_{\mu\alpha} B_{\alpha \nu \rho}+B_{\mu\alpha\rho}b_{\alpha\nu}]           &   -B_{\mu \alpha\rho} (G^{-1})^{\alpha \nu}  \\
 (G^{-1})^{\mu \alpha}  B_{\alpha\nu\rho}    &    0
\end{array}\right )\, ,
\end{equation}
are constant. We also introduced the double space vector
\begin{equation}\label{eq:dsv}
Z_{arg}^M = \left (
\begin{array}{c}
 V^\mu  \\
x^\mu
\end{array}\right )  =  \left (
\begin{array}{c}
-\kappa\theta_{0}^{\mu\nu} y_\nu+(g^{-1})^{\mu\nu} {\tilde{y}}_\nu  \\
x^\mu
\end{array}\right )\, ,
\end{equation}
according the rule that all background fields in the upper $D$ rows of (\ref{eq:gm}) depend on $V^\mu$ while
all background fields in the lower $D$ rows of (\ref{eq:gm}) depend on $ x^\mu$. For more details about notation see Section 4.
With the help of  (\ref{eq:vargdef}) we can conclude that $V^\mu$ and consequently   $Z_{arg}^M$    depend on both $y_\mu$ and its double $\tilde y_\mu$. It is important that we can not express
$Z_{arg}^M$ in terms of $Z^M$ because $\tilde y_\mu$ is not linear function on $y_\mu$. We can relate theirs derivatives as
\begin{equation}\label{eq:rzzarg}
\partial_\pm Z^M_{arg} = K_\pm {}^M {}_N \partial_\pm Z^N  \,  ,  \qquad
K_\pm {}^M {}_N  \equiv   \left (
\begin{array}{cc}
0        &   -\kappa \theta_{0 \pm}^{\mu \nu}  \\
 \delta^\mu_\nu     &   0
\end{array}\right )\, ,
\end{equation}
but the arguments of background fields does not appear with derivatives. This is significant differences in relation to a series of papers
\cite{HZ}-\cite{BT}, where the arguments of background fields depend on $Z^M$.

The finite part (the zeroth order) of the T-dual transformations (\ref{eq:xtdual}) have a form
\begin{equation}\label{eq:fp}
\partial_{\pm}x^\mu  \cong
-\kappa\theta^{\mu\nu}_{0\pm} \partial_{\pm} y_\nu \, ,  \qquad
\partial_{\pm}y_\mu  \cong
-2\Pi_{0 \mp\mu\nu} \partial_{\pm}x^\nu    \,  .
\end{equation}
The solution of these relation are
\begin{equation}\label{eq:soltd}
x^\mu \cong
-\kappa\,{\theta}_0^{\mu\nu} y_\nu
+(g^{-1})^{\mu\nu}\tilde{y}_\nu \,  , \qquad \quad
y_\mu \cong -2 b_{\mu \nu} x^\nu + G_{\mu \nu} \tilde{x}^\nu \,  .
\end{equation}
Note that solution for $x^\mu$ coincides with $V^\mu$ in (\ref{eq:vargdef}), so that we also can write
\begin{equation}\label{eq:xV}
 x^\mu \cong V^\mu  \,  .
\end{equation}

The generalized metric  satisfies the condition
\begin{equation}\label{eq:sonn}
(\Omega {\cal{H}})^2 =1 + 4 b h(x-V) \cong 1   \, ,
\end{equation}
because $x^\mu$ is T-dual to $V^\mu$.
This is the  consistency condition of the relation (\ref{eq:tdual}).
We can rewrite it as
\begin{equation}\label{eq:sodd}
{\cal{H}}^T  \Omega {\cal{H}} \cong \Omega \, ,
\end{equation}
because the generalized metric is symmetric,
${\cal{H}}^T \cong {\cal{H}}$.  As noticed in Refs.\cite{Duff,Hull}, the last relation  shows that there exists manifest $O(D,D)$ symmetry.

The inverse of the generalized metric has  a form
\begin{equation}
({\cal{H}}^{-1})^{MN}  = \left (
\begin{array}{cc}
  (G^{-1})^{\mu \rho} [A(x-V)]_\rho{}^\nu                  &         2 (G^{-1})^{\mu \rho} \, {}^{\diamond} B_{\rho \sigma}(V) [G^{-1} A(x-V) G]^\sigma{}_\nu  \\
 -2 \, {}^{\diamond} B_{\mu \rho}(x)  [G^{-1}A(x-V)]^{\rho \nu}       &    {}^{\diamond} G^E_{\mu \rho}(V) [G^{-1} A(x-V) G]^\rho{}_\nu
\end{array}\right )\, ,
\end{equation}
where $A(x)_\mu{}^\nu \equiv \delta_\mu{}^\nu - 6 (b G^{-1} h(x) G^{-1})_\mu{}^\nu $. In the zero order it takes the form
\begin{equation}
({\cal{H}}_{(0)}^{-1})^{MN}  = \left (
\begin{array}{cc}
  (G^{-1})^{\mu \nu}                  &         2 (G^{-1})^{\mu \rho} \, b_{\rho \nu}    \\
 -2 b_{\mu \rho}  (G^{-1})^{\rho \nu}       &    g_{\mu \nu}
\end{array}\right )\, .
\end{equation}


\subsection{Equations of motions as consistency condition of T-duality relations}

As  was discussed in  \cite{Duff,GR,DS1,ALLP}  the equation of motion and the Bianchi identity of the original theory are equal to the Bianchi identity and
the equation of motion of the T-dual theory. So, we will show that the consistency conditions of the relations (\ref{eq:tdual})
\begin{equation}\label{eq:tdual1}
\partial_+  [{\cal{H}}_{MN} (Z_{arg}) \partial_- Z^N] + \partial_-  [{\cal{H}}_{MN} (Z_{arg}) \partial_+ Z^N ] \cong 0  \, ,
\end{equation}
are T-dual to the equations of motion for both initial and T-dual theories.

We are going to multiply the last equation from the left with ${\cal{H}}^{-1}$. So, we will need expression
\begin{equation}\label{eq:hmjh}
{\cal{H}}^{-1} \partial_\pm {\cal{H}} = {\cal{H}}_0^{-1} \partial_\pm {\cal{H}} =  \partial_\pm
 \left (
\begin{array}{cc}
  -4 h(V) b +4 b h(x-V)                 &      -2 h(V)    \\
 2 [h(V)+4 b h(V) b] +2 [h(x-V)-4 b^2 h(x-V)]       &    4b h(V)
\end{array}\right )\, .
\end{equation}

Using the relation (\ref{eq:xV}),  $x^\mu \cong V^\mu$,  we have
\begin{equation}\label{eq:hmjh2}
{\cal{H}}^{-1} \partial_\pm {\cal{H}} \cong   \partial_\pm
 \left (
\begin{array}{cc}
  -4 h(x) b                &      -2 h(x)    \\
 2 [h(V)+4 b h(V) b]       &    4b h(V)
\end{array}\right )\, ,
\end{equation}
where in the first row we chose $x^\mu$ dependence and in the second row $V^\mu$ dependence.

Multiplying  (\ref{eq:tdual1}) from the left with ${\cal{H}}^{-1}$ and separating first and second rows
we obtain (for simplicity here we omit the indices)
\begin{eqnarray}\label{eq:tdualcom}
& 2 \partial_+ \partial_- x -6 \partial_+(hb) \partial_- x  -3 \partial_+ h \partial_- y    -6 \partial_- (hb) \partial_+ x
-3 \partial_- h \partial_+ y \cong 0 \,  , \nonumber \\
& 2 \partial_+ \partial_- y +3 \partial_+ (h+4bhb ) \partial_- x +6 \partial_+(bh) \partial_- y +  \nonumber \\
&  +3 \partial_- (h+4bhb ) \partial_+ x
+6 \partial_- (bh) \partial_+ y  \cong 0   \, ,
\end{eqnarray}
where all variables in the first equation depend on $x^\mu$ and in the second equation on $V^\mu$.

Using  he zeroth order T-dual transformations  (\ref{eq:fp}) these equations turn to
\begin{eqnarray}\label{eq:em}
 & \partial_+  \partial_- x^\mu - B^\mu{}_{\rho\sigma} \partial_+ x^\rho \partial_- x^\sigma \cong 0  \,  , \nonumber \\
 & \partial_+  \{ [ {}^\diamond \theta_{0-} (V) ]^{\mu \nu}   \partial_- y_\nu \}  -
 \partial_- \{ [{}^\diamond \theta_{0+} (V)]^{\mu \nu}   \partial_+ y_\nu \} \cong 0 \,  ,
\end{eqnarray}
where in analogy with (\ref{eq:risp}) we introduce
\begin{equation}\label{eq:dteta}
{}^\diamond \theta_{\pm} (V)  \equiv        \theta_{0 \pm} + \frac{3}{2} \Delta \theta_\pm (V)          \, ,
\end{equation}
with $\Delta \theta_\pm (V)= \theta_\pm (V) - \theta_{0 \pm} $.
So, they are just equations of motion for initial and T-dual theories, respectively.
Note that the second one can be rewritten as
\begin{eqnarray}
\partial_+ [\theta_{-}^{\mu \nu} (V)  \partial_- y_\nu -2 \theta_{0-}^{\mu \nu} \beta^+_\nu (V)]  -
\partial_- [\theta_{+}^{\mu \nu} (V)  \partial_+ y_\nu +2 \theta_{0+}^{\mu \nu} \beta^-_\nu (V)] \cong 0 \,  ,
\end{eqnarray}
which is  the form of the  equations of motion of T-dual theory from the Ref.\cite{DS1}.

The expression (\ref{eq:tdual1}) originated from conservation of the topological currents
$i^{\alpha M}= \varepsilon^{\alpha \beta} \partial_\beta Z^M$, which is often called Bianchi identity. So, we proved that T-duality transformations in the double space (\ref{eq:tdual}), for weakly curved background, unites equations of motion and Bianchi identities.

\cleq

\section{T-duality as coordinates permutations in flat double space}

The present article is generalization of the paper \cite{SB} for the case of weakly curved background. So, in this section we will repeat some notation and the results we are going to use.

Let us split  coordinate index $\mu$ into $a$ and $i$  ( $a=0,\cdots,d-1$,  $i=d,\cdots,D-1$),
and denote T-dualization along direction $x^a$ and $y_a$
\begin{equation}
{\cal T}^{a}=T^a \circ T_a  \,  ,  \quad    T^a \equiv T^{0}\circ T^{1}\circ\cdots\circ T^{d-1} \,  ,  \quad
T_a \equiv T_{0}\circ T_{1}\circ\cdots\circ T_{d-1} \,  .
\end{equation}

The main result of the paper \cite{SB} is the proof that exchange the places of some coordinates
$x^a$ with its T-dual $y_a$, in the flat double space produce the T-dual background fields
\begin{eqnarray}\label{eq:tdualf}
& {}_a \Pi_{0 \pm}^{ab} =   \frac{\kappa}{2} {\hat \theta}_{0 \mp}^{ab}  \, ,  \qquad \qquad        & {}_a \Pi_{0 \pm}^a{}_i =   \kappa {\hat \theta}_{0 \mp}^{ab} \Pi_{0 \pm bi}   \, ,   \\
& {}_a  \Pi_{0 \pm i}{}^a =  -\kappa \Pi_{0 \pm ib} {\hat \theta}_{0 \mp}^{ba}  \, ,       & {}_a \Pi_{0 \pm ij} = \Pi_{0 \pm ij} -2\kappa \Pi_{0 \pm ia} \hat\theta^{ab}_{0 \mp}\Pi_{0 \pm bj} \,  ,
\end{eqnarray}
where all notation are introduced in App. A.1.
The symmetric and antisymmetric parts of these expressions are T-dual metric and T-dual Kalb-Ramond field.
This is in complete agreement with the Ref.\cite{SN,DNS2}. The similar way to perform T-duality in the flat space-time for $D=3$ has been described in App. B
of Ref.\cite{ALLP}. Consequently, exchange the places of coordinates is equivalent to  T-dualization along
these coordinates.

As was shown in \cite{SB} eliminating $y_i$ from zero order T-duality transformations (\ref{eq:tdual})  gives
\begin{equation}\label{eq:pard0s}
\partial_\pm x^a \cong -2\kappa {\hat \theta}^{ab}_{0\pm} \Pi_{0\mp bi} \partial_\pm x^i - \kappa {\hat \theta}^{ab}_{0 \pm} \partial_\pm y_b \, .
\end{equation}
Similarly, eliminating $y_a$ from the same relation produces
\begin{equation}\label{eq:pard1s}
\partial_\pm x^i \cong -2\kappa {\hat \theta}^{ij}_{0\pm} \Pi_{0\mp ja} \partial_\pm x^a - \kappa {\hat \theta}^{ij}_{0 \pm} \partial_\pm y_j \, .
\end{equation}
The equation (\ref{eq:pard0s}) is zero order of the T-duality transformations for $x^a$ (eq. (44) of ref. \cite{DNS2}) and (\ref{eq:pard1s}) is its analogue
for $x^i$.

\cleq

\section{The complete T-duality chain in the weakly curved background}

Following the line of paper \cite{SB}  we will show that, in the case of the weakly curved background, the complete T-duality chain can be obtained
by by exchanging the places of coordinates in the double space. Due to the fact that background fields depend on the coordinates
this conjecture will be proven iteratively.

In the case of weakly curved background, comparing with the flat case one, the argument dependence is a new feature and will be discussed
in Subsections  4.1, 4.2 and 4.5.


\subsection{Notation for arguments of background fields}

In the flat space, permutation of the coordinates $x^a$ with the corresponding T-dual $y_a$, we realized by multiplying double space coordinate (\ref{eq:escoor})
\begin{equation}\label{eq:escoorai}
Z^M=\left (
\begin{array}{c}
 x^\mu  \\
 y_\mu
\end{array}\right )
=\left (
\begin{array}{c}
 x^a  \\
 x^i  \\
y_a  \\
y_i
\end{array}\right )\, ,
\end{equation}
by the  matrix
\begin{equation}\label{eq:taua}
({\cal T}^{a}) {}^M {}_N = \left (
\begin{array}{cc}
1-P_a  & P_a   \\
P_a  &  1- P_a
\end{array}\right ) =  \left (
\begin{array}{cccc}
0  & 0  & 1_a & 0  \\
0  & 1_i  & 0 & 0  \\
1_a & 0  & 0 & 0  \\
0  & 0  & 0 & 1_i
\end{array}\right )   =   1_2 \otimes (1-P_a) +   \Omega_{2} \otimes P_a     \,  .
\end{equation}
Here
\begin{equation}\label{eq:pata}
P_a = \left (
\begin{array}{cc}
1_a  &  0  \\
 0 &  0
\end{array}\right )\, ,
\end{equation}
is $D \times D$ projector with $d$ units on the main diagonal
where $1_a$ and $1_i$ are $d$ and $D-d$ dimensional identity matrices. It is easy to check that
\begin{equation}\label{eq:tprop}
({\cal T}^{a})^T = {\cal T}^{a}  \, , \qquad   ({\cal T}^{a}  {\cal T}^{a})^M {}_N  = \delta^M {}_N  \, , \qquad    (\Omega {\cal T}^{a} \Omega)^M {}_N  = ({\cal T}^{a})^M {}_N   \, , \qquad  {\cal T}^{a} \Omega  {\cal T}^{a}= \Omega \, .
\end{equation}

We have to be very careful with notation for arguments of background fields. The double space coordinate $Z^M$ has $2D$ rows. In the upper $D$ rows we have put the $D$ components   $x^\mu$ and in the lower $D$ rows the $D$ components  $y_\mu$.
The notation for  the arguments of background fields $Z_{arg} $  is a bit different.  Let us first notice that
in the same row of ${\cal{H}}(Z_{arg})$, as well as in the same row of combinations $ {\cal T}^{a} {\cal{H}}(Z_{arg}) \, {\cal T}^{a}$  (see (\ref{eq:dualh1})) the arguments of all background fields
are the same.  So, if we want to uniquely determine arguments of all background fields
it is enough to introduce new coordinate with $2D$ rows $Z_{arg} $, putting  in each row the corresponding argument from ${\cal{H}}(Z_{arg})$ or $ {\cal T}^{a} {\cal{H}}(Z_{arg}) \, {\cal T}^{a}$.
Note that unlike double space coordinate $Z^M$, where in each row there is only one component of the vector, in  arguments of background fields $Z_{arg} $ in each row there could be the complete $D$
dimensional vector. Rewritten in form of the one column the arguments of background fields are $2 D^2$ dimensional vector. In particular examples the background fields may depend on some $n \leq D$ component.
We will use indices $r,s$ to denote coordinates $x^r, \,\, (r=1,2, \cdots ,n)$, which appear as the background fields arguments.
Then the components of the argument of background fields are $n$-dimensional projections $x^r= P_r x^\mu$ and  $V^r= P_r V^\mu$ of $D$-dimensional vectors $x^\mu$ and  $V^\mu$.
In that case the arguments of background fields  will have  $2 D n$ nonzero elements and we will write them as $2 D n$ dimensional vector.

 We will use the following shorter notation. Let us start with the most complicated case for the  background fields combination  $ {\cal T}^{a} {\cal{H}}(Z_{arg}) \, {\cal T}^{a}$, which depend on $P_a$.
 The multiplication with matrices ${\cal T}^{a}$ partially implemented the change of arguments according to the relation (see (\ref{eq:dualh1}))
\begin{equation}\label{eq:pca}
 {\cal T}^{a} {\cal{H}}(Z_{arg}) \, {\cal T}^{a}  =  ( {\cal T}^{a} {\cal{H}} {\cal T}^{a}) (Z_{{\cal T}^{a} {\cal{H}} {\cal T}^{a}}) \, ,
\end{equation}
which can be used as definition of $Z_{{\cal T}^{a} {\cal{H}} {\cal T}^{a}}$.
It has   the same arguments in the first $d$ and then in the following  $D-d, \, d$ and $D-d$ rows, if we perform T-dualization along first $d$  rows. Generally, it has the same arguments in those $d$ rows of the first
$D$ rows defining by projection operator $P_a$ (which we will call $a$-rows) and in those $D-d$ rows of the first $D$ rows defining by projection operator $1- P_a$ (which we will call $i$-rows).
A similar rule  valid for the last $D$ rows. For simplicity, we will assume that we perform T-dualization along first $d$  rows, except in Sec.6   where we will present useful example.

So, in the shorter notation we will write such argument in four component notation (note that real dimension is $n [d+(D-d)+d+(D-d)]=2 D n$)
\begin{equation}\label{eq:bztht}
\breve Z_{{\cal T}^a {\cal{H}}  {\cal T}^a}
=\left |
\begin{array}{c}
 x^\mu    \\
V^\mu   \\
 V^\mu  \\
 x^\mu
\end{array}\right |_{a,r} =
\left |
\begin{array}{c}
 1   \\
0   \\
 0  \\
1
\end{array}\right|_a \otimes  x^r
+ \left |
\begin{array}{c}
0   \\
1  \\
1  \\
 0
\end{array}\right |_a \otimes  V^r
\, ,
\end{equation}
where the lower index  $a$  indicates the rule described above and index $r$  the projection with $P_r$.
With "breve" we mark $2D n$ dimensional vector.

Because arguments of all background fields in ${\cal{H}}(Z_{arg})$ and ${}_a {\cal{H}}(Z_{arg})$ are the same in the upper $D$ rows as well as in the lower $D$ rows we can
write them in two component notation. We will indicate it with index $D$. But, it is useful to reexpress these arguments  in $P_a$ dependent four component notation
\begin{equation}\label{eq:dsvp}
\breve Z_{arg}
=\left |
\begin{array}{c}
 V^\mu   \\
 x^\mu
\end{array}\right |_{D,r}
=\left |
\begin{array}{c}
V^\mu  \\
V^\mu   \\
 x^\mu   \\
 x^\mu
\end{array}\right |_{a,r} \, , \qquad
 {}_a \breve Z_{arg}
 =\left |
\begin{array}{c}
 x^a \, , V^i   \\
V^a \, ,  x^i
\end{array}\right |_{D,r}
=\left |
\begin{array}{c}
P_a \, x^\mu + (1-P_a)\, V^\mu  \\
P_a \, x^\mu + (1-P_a) \,V^\mu   \\
P_a \, V^\mu + (1-P_a) \,  x^\mu   \\
P_a \, V^\mu + (1-P_a) \,  x^\mu
\end{array}\right |_{a,r}
   \, ,
\end{equation}
in order to be of the same form as  $\breve Z_{{\cal T}^a {\cal{H}}  {\cal T}^a}$. Note that for example the first $D$ rows in ${}_a \breve Z_{arg}$ are
\begin{equation}
P_r [P_a \, x^\mu + (1-P_a)\, V^\mu]
=P_r \left (
\begin{array}{c}
 x^a    \\
V^i   \\
\end{array}\right ) \,
=\left |
\begin{array}{c}
 x^a    \\
V^i   \\
\end{array}\right|_{r} \, .
\end{equation}


\subsection{Relations between arguments of background fields}

The arguments (\ref{eq:bztht}) and (\ref{eq:dsvp})  are connected by the relations
\begin{equation}\label{eq:t2t1}
{}_a \breve Z_{arg} =     \breve{{\cal S}}^a  \,\, \breve Z_{{\cal T}^a {\cal{H}}  {\cal T}^a} \,  , \qquad  \breve Z_{{\cal T}^a {\cal{H}}  {\cal T}^a} =  \breve{{\cal R}}^a \, \, \breve Z_{arg}       \, ,
\end{equation}
where
\begin{eqnarray}\label{eq:R}
 \breve{{\cal R}}^a
 =  {\cal T}^a \otimes  P_r
=\left|
\begin{array}{cccc}
0 & 0 & 1_a P_r & 0\\
0 & 1_i P_r & 0 & 0\\
1_a P_r& 0 & 0 & 0 \\
0 & 0 & 0 & 1_i P_r
\end{array}
\right|_{a}   ,
\end{eqnarray}
and
\begin{eqnarray}\label{eq:S}
&  \breve{{\cal S}}^a
=  {\cal T}^a  \otimes (1 -P_a) P_r +  {\bar {\cal T}}^a \otimes P_a P_r        \nonumber \\         \nonumber \\
&=\left|
\begin{array}{cccc}
P_a P_r& 0 & (1-P_a) P_r& 0\\
0 & (1-P_a) P_r & 0 & P_a P_r \\
(1-P_a) P_r & 0 & P_a P_r & 0  \\
0 & P_a P_r & 0 & (1-P_a) P_r
\end{array}
\right|_{a}  \, .
\end{eqnarray}
The matrix ${\cal T}^a$  is defined in  (\ref{eq:taua})   and
\begin{eqnarray}
 {\bar {\cal T}}^a = 1_2 \otimes P_a +  \Omega_2  \otimes (1-P_a)
=\left(
\begin{array}{cccc}
1_a & 0 & 0 & 0\\
0 & 0 & 0 & 1_i \\
0 & 0 & 1_a & 0 \\
0 & 1_i & 0 & 0
\end{array}
\right)  \, .
\end{eqnarray}
The lower indices  $a$ and $r$ indicate the same  rules for the matrix as described above for vectors.
With "breve" here we marked $2D n \times 2D n$ matrices.

Because
\begin{equation}
({\cal T}^a)^2 = 1_{2D} \, \qquad   {\bar {\cal T}}^a  \, \cdot \,  {\cal T}^a = \Omega_{2D} \equiv \Omega_2 \otimes 1_D    \, ,
\end{equation}
we have
\begin{eqnarray}\label{eq:tsr}
& \breve{{\cal T}}^a =   \breve{{\cal S}}^a  \, \cdot \,  \breve{{\cal R}}^a = 1_{2} \otimes 1_D \otimes  (1-P_a) P_r  +  \Omega_{2} \otimes 1_D \otimes P_a P_r                  \nonumber \\  \nonumber \\
&= \left |
\begin{array}{cccc}
(1-P_a) P_r  & 0  &   P_a P_r &  0   \\
 0  &  (1-P_a) P_r  &  0 &  P_a P_r \\
P_a P_r & 0  &  (1- P_a) P_r  &  0   \\
 0  &  P_a P_r &  0 & (1- P_a) P_r
\end{array}\right|_{a}  \, .
\end{eqnarray}
According to  (\ref{eq:t2t1}) this matrix transforms the arguments of background fields
\begin{equation}\label{eq:bztz}
{}_a \breve Z_{arg} =     \breve{{\cal T}}^a  \, \, \breve Z_{arg}       \, .
\end{equation}
Note that double space coordinate $Z^M$ transforms with $2D \times 2D$ the matrix  (\ref{eq:taua}).
Up to $1_D$ (which reflects the fact that in any row we have put the $D$ dimensional vector) and $P_r$ (which, when it is different from one,  reflects the fact that we can omit background dependence of some coordinates),
${\cal T}^a$ and $ \breve{{\cal T}}^a$   have the same structure.
For $P_r=1$ they are isomorphic and  form the same group $G_T(D)$ with respect to multiplication. For $P_r \neq 1$ theirs groups are homeomorphic.   We can consider matrices ${\cal T}^a$ and $ \breve{{\cal T}}^a$ as different
representation of the same operator $\hat {\cal T}^a$.

The relation (\ref{eq:tsr}) shows the width of applicability of Buscher's approach. Whenever we perform T-dualization along coordinates $x^a$, which do not appear in the arguments of background fields, we will have
$P_a P_r =0$. Consequently in that case $\breve{{\cal T}}^a =1_{2D}\otimes P_r$ and  ${}_a \breve Z_{arg} = \breve Z_{arg} $. For $P_a P_r \neq  0$ (when we perform T-dualization along coordinates $x^a$, appearing in the arguments of background fields) the second term in (\ref{eq:tsr}) becomes nontrivial. It exchanges positions of $x^\mu$ with $V^\mu$ and turns the physical theory to the nonphysical, described  in Refs.\cite{DS1, DNS2}.

In order to get proper arguments   ${}_a \breve Z_{arg}$ from  arguments  $\breve Z_{{\cal T}^a {\cal{H}}  {\cal T}^a} $
 we should make substitution
\begin{equation}\label{eq:cof}
\begin{array}{c}
 x^\mu  \to  P_a \, x^\mu + (1-P_a)\, V^\mu     \quad \Longleftrightarrow    \quad     x^i \to  V^i               \\
 V^\mu   \to  P_a \, x^\mu + (1-P_a)\, V^\mu    \quad \Longleftrightarrow     \quad    V^a \to  x^a                     \\
 V^\mu  \to  P_a \, V^\mu + (1-P_a) \,  x^\mu   \quad \Longleftrightarrow     \quad    V^i \to  x^i                       \\
 x^\mu   \to P_a \, V^\mu + (1-P_a) \,  x^\mu   \quad \Longleftrightarrow     \quad    x^a \to  V^a
\end{array} \,  ,
\end{equation}
which performs the transformation   (\ref{eq:t2t1}) with the matrix $\breve {\cal S}^a$.
These rules are valid for $d, \, D-d, \, d, \,$ and $D-d$ rows of the generalized metric, respectively.
Consequently, in all background fields in ${\cal{T}}^a   {\cal{H}}  {\cal{T}}^a$ with left index $a$ (in all $a$ and $a+D$  rows) we should change all arguments with index $i: x^i \leftrightarrow   V^i$ in order to obtain ${}_a {\cal{H}}$.
Similarly,  in all background fields in ${\cal{T}}^a   {\cal{H}}  {\cal{T}}^a$ with left index $i$ (in all $i$ and $i+D$ rows) we should change all arguments with index $a: x^a \leftrightarrow   V^a$ in order to obtain ${}_a {\cal{H}}$.
 We can also obtain arguments  of ${}_a \breve Z_{arg}$   directly from $\breve Z_{arg} $ using   (\ref{eq:bztz}).


\subsection{T-duality transformations as coordinates permutations}

Let us derive expression for T-dual generalized metric.
We start with the T-duality relations in the form (\ref{eq:tdual})
\begin{equation}\label{eq:tdualwc}
\partial_{\pm} Z^M \cong \pm \, \Omega^{MN} {\cal{H}}_{NK} (Z_{arg}) \,\partial_{\pm}Z^K \, ,
\end{equation}
with generalized metric (\ref{eq:gm})
\begin{equation}\label{eq:gma}
{\cal{H}}_{MN} (Z_{arg}) = \left (
\begin{array}{cc}
{}^{\diamond} G^E_{\mu \nu}(V) &  -2 \, {}^{\diamond} B_{\mu\rho}(V) (G^{-1})^{\rho \nu}  \\
2 (G^{-1})^{\mu \rho} \, {}^{\diamond} B_{\rho \nu}(x) & (G^{-1})^{\mu \nu}
\end{array}\right )\, .
\end{equation}

We are going to apply the procedure of paper \cite{SB} to the case of weakly curved background. Then, beside double space coordinate $Z^M$  (\ref{eq:escoorai}) we should also transform
extended coordinates of the arguments of background fields $\breve Z_{arg}$  (\ref{eq:dsvp}).
We will require that the T-duality transformations (\ref{eq:tdualwc}) are invariant under transformations
of the  double space coordinates $Z^M$ and $\breve Z_{arg}$
\begin{equation}\label{eq:escoorait}
Z_a^M = {\cal T}^{a} {}^M {}_N Z^N  , \qquad
{}_a  \breve Z_{arg} =  \breve {\cal T}^{a} \breve Z_{arg} \, ,
\end{equation}
with the  matrices  ${\cal T}^{a}$ and $\breve {\cal T}^{a}$. The new coordinates should satisfy  the same form of  the equation
\begin{equation}\label{eq:tduald}
\partial_{\pm} Z_a^M \cong \pm \, \Omega^{MN} {}_a {\cal{H}}_{NK}({}_a Z_{arg}) \,\partial_{\pm}Z_a^K \, ,
\end{equation}
which produces the expression for the dual generalized metric in terms of the initial one
\begin{equation}\label{eq:dualgm}
{}_a {\cal{H}}({}_{a} Z_{arg} ) =  {\cal T}^{a} {\cal{H}}(Z_{arg}) \, {\cal T}^{a} \, .
\end{equation}
Note that  multiplication with matrices ${\cal T}^{a}$ partially implemented the change of arguments  according to (\ref{eq:pca}),
which we marked   with transformation matrix $\breve{\cal{R}}^a$
\begin{equation}\label{eq:dualgma}
{}_a {\cal{H}}({}_{a} Z_{arg} ) =  ({\cal T}^{a} \, {\cal{H}} \, {\cal T}^{a}) (\breve{\cal{R}}^a Z_{arg})\, .
\end{equation}
Recalling that ${}_{a} Z_{arg} = \breve {\cal T}^{a} Z_{arg} =  \breve{{\cal S}}^a \breve{{\cal R}}^a   Z_{arg} $ it is clear that in order to obtain ${}_{a} Z_{arg}$ we should make remaining transformation with
matrix $\breve{{\cal S}}^a$.

Consequently, we can obtain the dual generalized metric  ${}_a {\cal{H}}({}_{a} Z_{arg} )$ in two steps. First, multiplying initial one ${\cal{H}}(Z_{arg} )$  from  left and right with ${\cal T}^{a}$ and second additionally
transform the obtaining argument with matrix $\breve{{\cal S}}^a$.


\subsection{The T-dualities in the double space along all coordinates}

To learn what is going on with the arguments of background fields we will suppose that the first relation  (\ref{eq:escoorait}) valid and we will derive the second one
in the simplest case of complete T-dualization and when background fields depend on all the coordinates. In that case we have $P_a \to {}^\star P =1$ and $P_r \to 1$, so that (\ref{eq:taua})
turns to ${\cal T}^a  \to {}^\star {\cal T} = \Omega_2  \otimes 1_D =\left (
\begin{array}{cc}
0  &  1_D  \\
1_D  &  0
\end{array}\right )$
and the first relation  (\ref{eq:escoorait})   to  $Z_a^M  \to {}^\star Z^M = {}^\star {\cal T} Z^M =
\left (
\begin{array}{c}
 y_\mu  \\
 x^\mu
\end{array}\right )$.
Then from (\ref{eq:gma})  we obtain
\begin{equation}\label{eq:fdgm}
 {\cal T}^a  {\cal{H}} (Z_{arg} ) {\cal T}^{a} \to    {}^\star {\cal T}  {\cal{H}} (Z_{arg} ) {}^\star {\cal T}   = \left (
\begin{array}{cc}
 (G^{-1})^{\mu \nu}       &         2 (G^{-1})^{\mu \rho} \,   {}^{\diamond} B_{\rho \nu} (x)       \\
 -2  \,  {}^{\diamond} B_{\mu \rho}  (V) \, ( G^{-1})^{\rho \nu}       &     {}^{\diamond} G^E_{\mu \nu} (V)
\end{array}\right )\, .
\end{equation}
Using the analogy with known expressions for dual background fields  (\ref{eq:tdbf}), (\ref{eq:cdi1}) and (\ref{eq:cdi2}),
but this time according to (\ref{eq:ebf}) with diamond, we have
\begin{equation}\label{eq:fdgmd}
 {}^\star {\cal T}  {\cal{H}} (Z_{arg} ) {}^\star {\cal T}  = \left (
\begin{array}{cc}
 {}^{\diamond \star} G_E^{\mu \nu}  &  -2  ( {}^{\diamond \star} B \, {}^{\diamond \star} G^{-1})^\mu{}_\nu  (x) \\
2 ( {}^{\diamond \star} G^{-1}  \,   {}^{\diamond \star} B)_\mu {}^\nu (V)  & ( {}^{\diamond \star} G^{-1})_{\mu \nu} (V)
\end{array}\right )\, .
\end{equation}
Note that in the case of this subsection $\breve {\cal S}^{a} \to {}^\star \breve {\cal S} = {}^\star \bar {\cal T} = 1_{2D}$ and
$\breve{{\cal T}}^a \to     {}^\star \breve {\cal T} = {}^\star \breve {\cal R} = \Omega_2 \otimes 1_{D^2}$.
That is exactly what we expected, because  according to (\ref{eq:dualgma}) it should be  ${}^\star {\cal{H}}_{MN}  ( {}^\star Z_{arg})$. It  has the same form as the initial one (\ref{eq:gma})
but with T-dual background fields. We additionally learned that we also should exchange $x^\mu$ with $V^\mu$, because all background fields in upper $D$ rows depend on $x^\mu$ and
all background fields in lower $D$ rows depend on $V^\mu$.

So, as in the case of flat background   ${}^\star {\cal T}$ acts on $Z^M$,  but here ${}^\star \breve{{\cal T}}$   acts   on the arguments of background fields  $Z_{arg}$, as well. It exchanges $x^\mu$ with
$V^\mu$ as ${}^\star  Z_{arg} ={}^\star \breve{{\cal T}} Z_{arg}$ which is just the second eq.(\ref{eq:escoorait}) for $P_a \to {}^\star P=1\, , P_r \to 1$, or explicitly
\begin{equation}\label{eq:tdarg}
 \begin{array}{c}
{}^\star  \\
 .
\end{array}
\left |
\begin{array}{c}
 x^\mu  \\
 x^\mu  \\
V^\mu  \\
V^\mu
\end{array}\right|_a  =\left |
\begin{array}{cccc}
0 & 0 & 1 & 0 \\
0 & 0 & 0 & 1 \\
1 & 0 & 0 & 0 \\
0 & 1 & 0 & 0
\end{array}\right |_a
\left |
\begin{array}{c}
V^\mu  \\
V^\mu  \\
x^\mu  \\
x^\mu
\end{array}\right |_a   \, .
\end{equation}
So, we should transform both $Z^M$ and $Z_{arg}$ with the matrices  ${}^\star {\cal T}= \Omega_2 \otimes 1_D$ and ${}^\star \breve {\cal T} = \Omega_2 \otimes 1_{D^2}$, in accordance with the (\ref{eq:escoorait}).


\subsection{The arguments of the dual background fields}

Up to now the variable $V^\mu$ was undetermined.  Now we will determine it for for arbitrary T-dualization.
Because, all arguments are multiplied by infinitesimal coefficient $B_{\mu \nu \rho}$
we can calculate them using the zero order of the T-duality transformations (\ref{eq:tdual}).
So, the solution for $x^a$ we proclaim as $V^a$ and from (\ref{eq:pard0s}) we obtain
\begin{equation}\label{eq:va}
\partial_\pm V^a = -2\kappa {\hat \theta}^{ab}_{0\pm} \Pi_{0\mp bi} \partial_\pm x^i - \kappa {\hat \theta}^{ab}_{0 \pm} \partial_\pm y_b \, ,
\end{equation}
which coincide with eq.(37) of Ref.\cite{DNS2}.

Similarly, we proclaim solution for $x^i$ as $V^i$ and from (\ref{eq:pard1s}) we obtain
\begin{equation}\label{eq:vi}
\partial_\pm V^i = -2\kappa {\hat \theta}^{ij}_{0\pm} \Pi_{0\mp ja} \partial_\pm x^a - \kappa {\hat \theta}^{ij}_{0 \pm} \partial_\pm y_j \, .
\end{equation}
This is in fact the same relation (\ref{eq:va}) with altered indices $i,j \leftrightarrow a,b$.

So far we obtained the expressions for $V^a$ and $V^i$. Now, we will determine theirs positions.
The multiplication with matrices ${\cal T}^{a}$ partially implemented the change of arguments according to the relation  (\ref{eq:pca})
\begin{eqnarray}\label{eq:dualh1}
&     {\cal T}^{a} {\cal{H}}(Z_{arg}) \, {\cal T}^{a}  =  ( {\cal T}^{a} {\cal{H}} {\cal T}^{a}) (Z_{{\cal T}^{a} {\cal{H}} {\cal T}^{a}}) \,     \nonumber \\           \nonumber \\
&   = \left (
\begin{array}{cccc}
(G^{-1})^{ab}  (x)     & 2 (G^{-1}{}^{\diamond} B)^a{}_j (x) &   2 (G^{-1} {}^{\diamond} B)^a{}_b  (x) & (G^{-1})^{aj} (x) \\
-2 ({}^{\diamond} B G^{-1})_i{}^b   (V)  & {}^{\diamond} G^E_{ij}    (V)         &  {}^{\diamond} G^E_{ib}    (V)           &  -2 ({}^{\diamond} B G^{-1})_i{}^j  (V)\\
-2 ( {}^{\diamond} B G^{-1})_a{}^b (V)   & {}^{\diamond} G^E_{aj}  (V)         &   {}^{\diamond} G^E_{ab}   (V)     & -2 ({}^{\diamond} B G^{-1})_a{}^j  (V) \\
(G^{-1})^{ib} (x)      & 2 (G^{-1} {}^{\diamond} B)^i{}_j (x) & 2 (G^{-1} {}^{\diamond} B)^i{}_b   (x)  & (G^{-1})^{ij} (x)
\end{array}\right )\, .
\end{eqnarray}

To obtain final result, after transformation with matrix $\breve{{\cal S}^{a}}$   we have
\begin{eqnarray}\label{eq:dualh2}
& {}_a {\cal{H}} ({}_a Z_{arg})   =  \nonumber \\ \nonumber \\
&= \left (
\begin{array}{cccc}
(G^{-1})^{ab}  (x^a,V^i)     & 2 (G^{-1}{}^{\diamond} B)^a{}_j (x^a,V^i)  &   2 (G^{-1} {}^{\diamond} B)^a{}_b  (x^a,V^i)  & (G^{-1})^{aj} (x^a,V^i) (x^a,V^i)  \\
-2 ({}^{\diamond} B G^{-1})_i{}^b   (x^a,V^i)   & {}^{\diamond} G^E_{ij}   (x^a,V^i)        &  {}^{\diamond} G^E_{ib}  (x^a,V^i)       &  -2 ({}^{\diamond} B G^{-1})_i{}^j  (x^a,V^i) \\
-2 ( {}^{\diamond} B G^{-1})_a{}^b (V^a,x^i)    & {}^{\diamond} G^E_{aj}  (V^a,x^i)         &   {}^{\diamond} G^E_{ab} (V^a,x^i)     & -2 ({}^{\diamond} B G^{-1})_a{}^j  (V^a,x^i) \\
(G^{-1})^{ib} (V^a,x^i)      & 2 (G^{-1} {}^{\diamond} B)^i{}_j (V^a,x^i) & 2 (G^{-1} {}^{\diamond} B)^i{}_b  (V^a,x^i)  & (G^{-1})^{ij} (V^a,x^i)
\end{array}\right )\, ,
\end{eqnarray}
where $V^a= V^a(x^i, y_a)$  and $V^i= V^i(x^a, y_i)$  are defined in (\ref{eq:va}) and (\ref{eq:vi}) respectively.


\subsection{The T-dual background fields in the weakly curved background}

As well as in the case of the flat background we will require that the dual generalized metric  has the same form as the initial one
(\ref{eq:gm})   but with T-dual background fields
\begin{equation}\label{eq:hd}
{}_a {\cal{H}}_{MN} ({}_a Z_{arg}) = \left (
\begin{array}{cc}
{}_a  {}^{\diamond} G_E^{\mu \nu} (x^a, V^i) &  -2  ({}_a  {}^{\diamond} B \,{}_a {}^{\diamond} G^{-1})^\mu{}_\nu  (x^a, V^i) \\
2 ({}_a {}^{\diamond} G^{-1}\, {}_a  {}^{\diamond} B)_\mu {}^\nu (V^a, x^i)  & ({}_a {}^{\diamond} G^{-1})_{\mu \nu} (V^a, x^i)
\end{array}\right )\, .
\end{equation}
From now on, in this subsection, we will omit the arguments of background fields, because  all fields in lower $D$ rows depend on the same variables $(V^a, x^i)$.

It is useful to consider background fields
\begin{equation}\label{eq:dppmwc1}
{}_a {}^{\diamond}  \Pi_{ \pm}^{\mu \nu}  \equiv ({}_a {}^{\diamond}   B  \pm \frac{1}{2} {}_a {}^{\diamond}  G )^{\mu \nu} =   {}_a {}^{\diamond}  G^{\mu \rho}
 [({}_a {}^{\diamond} G^{-1} \, {}_a {}^{\diamond}   B )_\rho {}^\nu \pm \frac{1}{2} \delta^\nu_\rho ] \,  \, ,
\end{equation}
and express them in terms of initial fields. From lower $D$ rows of expressions (\ref{eq:dualh2}) and (\ref{eq:hd}) we have
\begin{equation}\label{eq:bgd}
( {}_a {}^{\diamond} G^{-1} \, {}_a {}^{\diamond} B)_\mu{}^\nu  = \left (
\begin{array}{cc}
- ({}^{\diamond} B  G^{-1})_a{}^b       & \frac{1}{2} {}^{\diamond}  G^E_{aj}    \\
  \frac{1}{2} ( G^{-1})^{ib}    &   ( G^{-1} {}^{\diamond}  B)^i{}_j
\end{array}\right )  \equiv \left (
\begin{array}{cc}
- {}^{\diamond} {\tilde \beta}    &  \frac{1}{2} {}^{\diamond} G_E^T   \\
 \frac{1}{2}  \gamma  &   - {}^{\diamond} {\bar \beta}^T
\end{array}\right )              \,  ,
\end{equation}
and
\begin{equation}\label{eq:gmjdi}
({}_a {}^{\diamond} G^{-1})_{\mu \nu}  = \left (
\begin{array}{cc}
{}^{\diamond} G^E_{ab}              & -2 ({}^{\diamond} B G^{-1})_a{}^j  \\
2 (G^{-1} {}^{\diamond} B)^i{}_b     & (G^{-1})^{ij}
\end{array}\right )  \equiv \left (
\begin{array}{cc}
{}^{\diamond} {\tilde G}_E   &   -2 \, {}^{\diamond} \beta_1 \\
-2 \, {}^{\diamond} \beta_1^T    &   {\bar \gamma}
\end{array}\right )     \, ,
\end{equation}
where we used  (\ref{eq:gnmj}), (\ref{eq:dgm}), (\ref{eq:dbgnmj}) and (\ref{eq:gnmjdb})  to obtain the second equalities.
To calculate the inverse of the last expression we will use the general expression for block wise inversion matrices
\begin{equation}\label{eq:im}
\left(\begin{array}{cc}
A & B\\
C & D
\end{array}\right)^{-1}\,=
\left(
\begin{array}{cc}
(A-BD^{-1}C)^{-1}          &     -A^{-1}B(D-CA^{-1}B)^{-1}\\
-D^{-1}C(A-BD^{-1}C)^{-1}  &      (D-CA^{-1}B)^{-1}
\end{array}\right).
\nonumber\\
\end{equation}
So, we find that
\begin{equation}\label{eq:gmjd}
({}_a {}^{\diamond} G)^{\mu \nu}  = \left (
\begin{array}{cc}
 ({}^{\diamond} A^{-1})^{ab}                              &   2 ({}^{\diamond} {\tilde G}_E^{-1} {}^{\diamond} \beta_1 {}^{\diamond} D^{-1})^a{}_{j} \\
2 ({\bar \gamma}^{-1} {}^{\diamond} \beta_1^T  {}^{\diamond} A^{-1})_i{}^{b}                &  ( {}^{\diamond} D^{-1})_{ij}
\end{array}\right )\, ,
\end{equation}
where we introduced
\begin{equation}\label{eq:AD1}
{}^{\diamond} A_{ab}= ({}^{\diamond} {\tilde G}_E -4 \, {}^{\diamond} \beta_1  {\bar \gamma}^{-1} {}^{\diamond} \beta_1^T)_{ab}  \, ,  \qquad \quad
{}^{\diamond} D^{ij} = ({\bar \gamma} -4 \, {}^{\diamond} \beta_1^T {}^{\diamond} {\tilde G}_E^{-1} {}^{\diamond} \beta_1)^{ij}  \, .
\end{equation}

It can be shown that
\begin{equation}\label{eq:A1}
{}^{\diamond} A_{ab}= ( {\tilde G} -4 \, {}^{\diamond}  {\tilde B}  {\tilde G}^{-1} \, {}^{\diamond} {\tilde B})_{ab} \equiv {}^{\diamond} \hat{G}^E_{ab}  \, ,
\end{equation}
where the last expression is definition (\ref{eq:ghatwc}).
Similarly as in the flat space case, $ {}^{\diamond} {\tilde G}^E_{ab}$, is just $ab$ component of ${}^{\diamond} G^E_{\mu \nu}$,
while the ${}^{\diamond} \hat{G}^E_{ab}$ has the same form as effective metric $G^E_{\mu \nu}$ but with all components (${\tilde G}, {}^{\diamond} {\tilde B}$)
defined in $d$ dimensional subspace with indices $a,b$.

With the help of (\ref{eq:A1}) we can put the first equation (\ref{eq:AD1}) in the form
$({}^{\diamond} \hat{G}^E)_{ab}  = ({}^{\diamond} {\tilde G}^E)_{ab}  -4 ({}^{\diamond} \beta_1  {\bar \gamma}^{-1} {}^{\diamond} \beta_1^T)_{ab}$.
Multiplying it on the left with $({{}^{\diamond} {\tilde G}_E}^{-1})^{ab}$ and on the right with $({{}^{\diamond} {\hat G}_E}^{-1})^{ab}$ we obtain
\begin{equation}\label{eq:orwc}
({{}^{\diamond} {\tilde G}_E}^{-1})^{ab} = ({{}^{\diamond} {\hat G}_E}^{-1})^{ab} -4 \,
 ({{}^{\diamond} {\tilde G}_E}^{-1} {}^{\diamond} \beta_1  {\bar \gamma}^{-1} {}^{\diamond} \beta_1^T {{}^{\diamond} {\hat G}_E}^{-1})^{ab}  \, ,
\end{equation}
which help us to verify that
\begin{equation}\label{eq:Dinv}
({}^{\diamond}  D^{-1})_{ij} = (\bar{\gamma}^{-1} + 4 \bar{\gamma}^{-1} {}^{\diamond}  \beta_1^T  {{}^{\diamond} {\hat G}_E}^{-1} {}^{\diamond} \beta_1  \bar{\gamma}^{-1})_{ij} \, ,
\end{equation}
is inverse of the second equation (\ref{eq:AD1}).

Using (\ref{eq:bgd}) and (\ref{eq:gmjd})  we can express background field (\ref{eq:dppmwc1}) as
\begin{equation}\label{eq:dppmwc12}
{}_a {}^{\diamond} \Pi_{\pm}^{\mu \nu}  = \left (
\begin{array}{cc}
{}^{\diamond} {\tilde G_E}^{-1} {}^{\diamond}\beta_1 {}^{\diamond}D^{-1} \gamma - {}^{\diamond}A^{-1} ({}^{\diamond}{\tilde \beta} \mp \frac{1}{2})         &   \frac{1}{2} {}^{\diamond}A^{-1} {}^{\diamond} G_E^T -2 \, {}^{\diamond}{\tilde  G}_E^{-1} {}^{\diamond}\beta_1 {}^{\diamond}D^{-1} ({}^{\diamond}{\bar \beta}^T \mp \frac{1}{2})    \\
 \frac{1}{2}{}^{\diamond}D^{-1} \gamma -2 \bar{\gamma}^{-1} {}^{\diamond}\beta_1^T {}^{\diamond}A^{-1} ({}^{\diamond}{\tilde \beta} \mp \frac{1}{2})      &   \bar{\gamma}^{-1} {}^{\diamond}\beta_1^T {}^{\diamond}A^{-1} {}^{\diamond} G_E^T - {}^{\diamond}D^{-1} ({}^{\diamond}{\bar \beta}^T \mp \frac{1}{2})
\end{array}\right )\, .
\end{equation}
After lengthy computation using  (\ref{eq:cgama}), (\ref{eq:dbgnmj}) and (\ref{eq:gnmjdb})  we find
\begin{equation}\label{eq:pipmr}
{}_a {}^{\diamond} \Pi_{ \pm}^{\mu \nu}  = \left (
\begin{array}{cc}
 \frac{\kappa}{2} {}^{\diamond} {\hat \theta}_{ \mp}^{ab}   &  \kappa {}^{\diamond} {\hat \theta}_{\mp}^{ab} {}^{\diamond} \Pi_{\pm bi}  \\
-\kappa {}^{\diamond} \Pi_{\pm ib} {}^{\diamond} {\hat \theta}_{ \mp}^{ba}    &  {}^{\diamond} \Pi_{\pm ij} -2\kappa {}^{\diamond}\Pi_{\pm ia} {}^{\diamond} \hat\theta^{ab}_{\mp} {}^{\diamond} \Pi_{\pm bj}
\end{array}\right )\, ,
\end{equation}
where ${}^{\diamond} \Pi_{\pm ab}$ and  ${}^{\diamond} {\hat \theta}_\pm^{ab}$ are defined in (\ref{eq:bfcwc}).

Comparison of the upper $D$ rows of expressions (\ref{eq:dualh1}) and (\ref{eq:hd}) yields the same result.
In component notation we can write
\begin{eqnarray}\label{eq:tdualf}
& {}_a {}^{\diamond} \Pi_{ \pm}^{ab} =   \frac{\kappa}{2} {}^{\diamond} {\hat \theta}_{ \mp}^{ab}  \, ,  \qquad \qquad        & {}_a {}^{\diamond}\Pi_{ \pm}^a{}_i =   \kappa {}^{\diamond} {\hat \theta}_{ \mp}^{ab} {}^{\diamond} \Pi_{ \pm bi}   \, ,   \nonumber \\
& {}_a {}^{\diamond}  \Pi_{ \pm i}{}^a =  -\kappa {}^{\diamond} \Pi_{ \pm ib} {}^{\diamond} {\hat \theta}_{ \mp}^{ba}  \, ,       & {}_a {}^{\diamond} \Pi_{ \pm ij} = {}^{\diamond} \Pi_{ \pm ij} -2\kappa {}^{\diamond} \Pi_{ \pm ia} {}^{\diamond} \hat\theta^{ab}_{ \mp} {}^{\diamond} \Pi_{ \pm bj} \,  .
\end{eqnarray}
The equality of two redefined background fields with diamond means that both finite and infinitesimal parts are equal.
According to (\ref{eq:ebf}) it follows that
the same relations are valid for  background fields without diamond
\begin{eqnarray}\label{eq:tdualf}
& {}_a \Pi_{ \pm}^{ab} =   \frac{\kappa}{2}  {\hat \theta}_{ \mp}^{ab}  \, ,  \qquad \qquad        & {}_a \Pi_{ \pm}^a{}_i =   \kappa  {\hat \theta}_{ \mp}^{ab}  \Pi_{ \pm bi}   \, ,  \nonumber  \\
& {}_a  \Pi_{ \pm i}{}^a =  -\kappa \Pi_{ \pm ib}  {\hat \theta}_{ \mp}^{ba}  \, ,       & {}_a \Pi_{ \pm ij} =  \Pi_{ \pm ij} -2\kappa  \Pi_{ \pm ia}  \hat\theta^{ab}_{ \mp} \Pi_{ \pm bj} \,  .
\end{eqnarray}
The arguments of all background fields are $[V^a (x^i,y_a),x^i]$, where $V^a$ is defined in (\ref{eq:va}).

Consequently, we obtained the T-dual background fields in the weakly curved background after dualization along directions $x^a \, (a=0,1,\cdots ,d-1)$.
They are in complete agreement with eq.(42)  of Ref.\cite{DNS2}.
So, the generalized metric ${\cal{H}}_{MN} (Z_{arg})$ contains sufficient information about background fields in each node of the chain (\ref{eq:chain2}).

The  double space contains coordinates of two spaces  totally dual relative to one another.
The starting theories are: the initial one described by the action $S(x^\mu)$ and its T-dual along all coordinates with the action  $S(y_\mu)$. Arbitrary T-dualization ${\cal T}^{a}$,
in the double space along $d$ coordinate with index $a$,  transforms at the same time $S(x^\mu)$ to $S[V^a(x^i, y_a),x^i]$ and
$S(y_\mu)$ to $S[x^a, V^i(x^a, y_i)]$. The obtained theories are also totally dual relative to one another.


\subsection{T-duality group $G_T(D)$,  for the weakly curved background}

Although, in the weakly curved background not only the double coordinate $Z^M$ but also the argument of background field $\breve Z_{arg}$ should be transformed, they are transformed  with the same operator
$\hat {\cal T}^{a}$. Because they live in the spaces with different dimensions, the operator $\hat {\cal T}^{a}$  has different representations  ${\cal T}^{a}$ and   $\breve {\cal T}^{a}$.

Successively T-dualization  can be represent by operator  multiplications $\hat {\cal T}^{a_1} \hat {\cal T}^{a_2} =  \hat {\cal T}^{a}, \,\, (a= a_1 \bigcup a_2)$.
The sets of operators  $\hat {\cal T}^{a}$ form a  commutative group $G_T(D)$ with respect to  multiplication.
Consequently, the set of all T-duality transformations form the group  $G_T(D)$ with respect to the operation $\circ $.

This is a subgroup of the $2D$ permutational group and T-duality along $2d$ coordinates $x^a \,$ and  $y_a \,$  can be represent as
\begin{equation}\label{eq:pg}
 \left (
\begin{array}{ccccccccccccc}
1      & 2   & \cdots    &   d  & d+1 & \cdots & D & D+1 & \cdots &   D+d  & D+d+1 & \cdots & 2D \\
D+1   &  D+2  & \cdots  &  D+d & d+1  &\cdots & D & 1      & \cdots & d  & D+d+1 & \cdots & 2D
\end{array}\right )\, ,
\end{equation}
or in the cyclic notation
\begin{equation}\label{eq:cpg}
(1,D+1) (2,D+2) \cdots (d, D+d)   \, .
\end{equation}
So, T-duality group is the same for the flat and weakly curved background. It is a global symmetry group of equations of motion (\ref{eq:tdual1}).


\cleq

\section{Dilaton}

To include dilaton field $\phi$,  we will add Fradkin-Tseytlin term to the action (\ref{eq:action0})
\begin{equation}\label{eq:FTa}
S_\phi = \int d^2 \xi \sqrt{-g} R^{(2)} \phi  \, ,
\end{equation}
where $R^{(2)}$ is scalar curvature of the world sheet. It is one order higher in $\alpha^\prime$ then terms in (\ref{eq:action0}) with metric and B-field.

Let us consider the  T-duality transformation of the dilaton field in the weakly curved background. There are several topics we are going to discuss.

\subsection{Path integral in the weakly curved background}

It is well known that dilaton transformation has quantum origin. For a constant background the Gaussian path integral produces
the expression $(\det \Pi_{0 + \mu \nu})^{-1}$. In our case the background is coordinate dependent and the integration is not Gaussian. The fact that we have infinitesimally small parameter will help us to show that the final result is formally the same as in the flat case.  Let us start with path integral
\begin{equation}\label{eq:fi}
{\cal Z} = \int d v_+^\mu   d v_-^\mu dy_\mu    e^{i S_{fix} (v_{\pm}, \partial_\pm y)}   \, ,
\end{equation}
where
\begin{eqnarray}
&    S_{fix} (v_{\pm}, \partial_\pm y) = S_0 + S_1 \,  ,                \nonumber \\
 \nonumber \\
& S_0 = \kappa \int  d^2 \xi [v_+ \Pi_{0+}  v_- + \frac{1}{2} (v_+ \partial_- y - v_- \partial_+ y ) ] \,  , \qquad   S_1 = \kappa \int  d^2 \xi \, v_+ h(V) v_-   \,  ,
\end{eqnarray}
and $S_1$ is infinitesimal. For simplicity, in this Subsection  we will omit space-time indices.  We will consider $\partial_\mp y$ as a sources for $v_\pm$. We introduce differential operators
\begin{equation}\label{eq:do}
\hat v_\pm = \mp  \frac{2i}{\kappa} \frac{\delta}{\delta \partial_\mp y}   \, ,  \qquad  \hat V = \int (d \xi^+ \hat v_+ + d \xi^- \hat v_- ) \, ,
\end{equation}
such that
\begin{equation}\label{eq:odo}
\hat v_\pm   e^{i S_{0} (v_{\pm}, \partial_\pm y)}  = v_\pm   e^{i S_{0} (v_{\pm}, \partial_\pm y)} \,  ,\qquad
\hat V   e^{i S_{0} (v_{\pm}, \partial_\pm y)}  = V   e^{i S_{0} (v_{\pm}, \partial_\pm y)} \, .
\end{equation}
Using the fact  that $S_1$ is infinitesimal we can write
\begin{equation}\label{eq:fi1}
{\cal Z} = \int dy_\mu   [1+ i \kappa \int d^2 \xi \, \hat v_+ h(\hat V)  \hat v_- ]   \, \int d v_+^\mu   d v_-^\mu     e^{i S_{0} (v_{\pm}, \partial_\pm y)}   \, .
\end{equation}
The integral over $v_\pm$ is Gaussian and we can  find
\begin{equation}\label{eq:fi2}
{\cal Z} = \int dy_\mu   \left[ 1+ i \kappa \int d^2 \xi (\hat v_+ h(\hat V)  \hat v_-) \right]   \, \frac{1}{det(\Pi_{0+})}  e^{i \, {}^\star S_{0} (y)}   \, ,
\end{equation}
where ${}^\star S_{0} ( y)  = \frac{\kappa^2}{2} \int d^2 \xi \, \partial_+ y \, \theta_{0-} \partial_- y$ is zero order of the T-dual action. Because
the determinant is constant  we should apply  differentiation only on ${}^\star S_{0} (y)$. After some direct calculation we obtain
\begin{equation}\label{eq:difs0}
 i \kappa \int d^2 \xi (\hat v_+ h(\hat V)  \hat v_-)    \,  e^{i \, {}^\star S_{0} (y)}  =
 \left[ i \frac{\kappa^2}{2} \int d^2 \xi \, \partial_+ y \, \theta_{1-}(V) \partial_- y  +  2\kappa \theta_{0-}^{\mu \nu} \int d^2 \xi \, h_{\mu \nu} (V) \right]
 e^{i \, {}^\star S_{0} (y)} \, ,
\end{equation}
where $\theta_{1-}(V) = -2\kappa \theta_{0-}  h (V) \theta_{0-}$. Substitute it back into (\ref{eq:fi2}) and express the infinitesimal term as an exponent we have
\begin{equation}\label{eq:fi3}
{\cal Z} = \int dy_\mu    \, \frac{1}{det(\Pi_{0+})}   e^{ 2\kappa \theta_{0-}^{\mu \nu} \int h_{\mu \nu}(V)}  \, \cdot \,   e^{i \, {}^\star S (y) }   \, ,
\end{equation}
where ${}^\star S (y)= \frac{\kappa^2}{2} \int d^2 \xi \, \partial_+ y \, [\theta_{0-} + \theta_{1-}(V) ] \partial_- y$ is the full T-dual action.

In the first two terms we can recognize $det(\Pi_{+}(V))$. Separating the infinitesimal term we have
$\Pi_{+}(V)= \Pi_{0+}+ h(V)$ and consequently
\begin{equation}\label{eq:piab}
\det \Pi_{+}(V)= \det \Pi_{0+} \det (1+ 2\kappa {\theta}_{0-} h )   \, .
\end{equation}
Using the fact that $h$ is infinitesimal we have, up to the first order
\begin{equation}\label{eq:}
\det (1+ 2\kappa { \theta}_{0-} h ) = 1+ Tr ( 2\kappa { \theta}_{0-} h) = e^{Tr ( 2\kappa {\theta}_{0-} h)}  = e^{- 2\kappa { \theta}_{0-}^{\mu \nu} \int h_{\mu \nu}} \, .
\end{equation}
From last three equations we obtain
\begin{equation}\label{eq:fi4}
{\cal Z} = \int dy_\mu    \, \frac{1}{det(\Pi_{+}(V))}   \,   e^{i \, {}^\star S (y) }   \, .
\end{equation}
Consequently, the functional integration over $v_\pm$ in non-Gaussian case for weakly curved background (where the action is of the third degree)  produces formally the same result as in the flat space
(where the action is Gaussian).


\subsection{Functional measure in the weakly curved background}

Using the expressions for T-dual fields  (\ref{eq:tdualf})   we can find the relations between the determinants
\begin{equation}\label{eq:dr}
\det (2 \Pi_{\pm ab})=  \frac{1}{\det ( 2 \, {}_a \Pi_{\pm}^{ ab})} = \sqrt{\frac{\det G_{a b}}{\det {}_a G^{a b}}}   \, .
\end{equation}
We put factor $2$ for the convenience, because $\Pi_{\pm ab} = B_{a b}\pm \frac{1}{2}G_{a b}$.
On the other hand, from   (\ref{eq:G}) and (\ref{eq:gmjdi})   we have
\begin{equation}\label{eq:drg}
\det G_{\mu \nu}=  \frac{\det G_{a b}}{\det {\bar \gamma}^{i j}} \, , \qquad  \det {}_a G_{\mu \nu}=  \frac{\det {}_a G^{a b}}{\det {\bar \gamma}^{i j}}  \, .
\end{equation}
From  (\ref{eq:dr}) and (\ref{eq:drg})  follows
\begin{equation}\label{eq:drf}
\det (2 \Pi_{\pm ab})=  \frac{1}{\det ( 2 \, {}_a \Pi_{\pm}^{ ab})} = \sqrt{\frac{\det G_{\mu \nu}}{\det {}_a G_{\mu \nu}}}   \, .
\end{equation}

With the help of last relation we can show that the change of space-time measure in the path integral of the ordinary (not doubled) approach  is correct
\begin{equation}\label{eq:ccm}
\sqrt{\det G_{\mu \nu}} \, d x^i d x^a  \to     \sqrt{\det G_{\mu \nu}}\,  d x^i    \frac{1}{\det ( 2 \,  \Pi_{+ ab})}\,  d y_a =
\sqrt{\det {}_a G_{\mu \nu}} \, d x^i d y_a \, ,
\end{equation}
when we performed T-dualization $T^a$ along  $x^a$ directions.

In the double space the  path integral measure is invariant under T-dual transformation. In fact after T-dualization ${\cal T}^a$ along both $x^a$ and $y_a$ we have
\begin{eqnarray}\label{eq:dim}
& \sqrt{\det G_{\mu \nu}} \,    \sqrt{\det {}^\star G^{\mu \nu}} \,d x^i    d y_i d x^a d y_a   \to   \\ \nonumber
 \\ \nonumber
& \sqrt{\det G_{\mu \nu}} \,   \sqrt{\det {}^\star G^{\mu \nu}} \,d x^i    d y_i  d y_a  d x^a
\frac{1}{\det ( 2 \,  \Pi_{+ ab}) \det ( 2 \, {}_a \Pi_{+}^{ ab})}\,    \,  ,
\end{eqnarray}
and because according to (\ref{eq:dr}) the last term is equal to $1$ we can conclude that the measure is invariant.


\subsection{Dilaton in the double space}

In double space the new dilaton should be introduced, invariant under T-duality transformations.
As it is explained in Refs.\cite{B,GR} the expression $\det ( 2 \Pi_{\pm ab})$ produces the well known  shift in the dilaton transformation
\begin{equation}\label{eq:dsh}
 {}_a \phi  = \phi - \ln \det ( 2 \Pi_{+ ab}) = \phi -  \ln \sqrt{\frac{\det G_{a b}}{\det {}_a G^{a b}}} \, .
\end{equation}
Note that according to (\ref{eq:dr}) we have
\begin{equation}\label{eq:invd}
 {}_a ({}_a \phi)  =  {}_a \phi - \ln \det ( 2  {}_a \Pi_{+ ab})  = \phi - \ln \det ( 2 \Pi_{+ ab})  - \ln \det \frac{1}{( 2 \Pi_{+ ab})} = \phi  \, ,
\end{equation}
which means that $\phi + {}_a \phi$ is duality invariant. So, we can define  dilaton
\begin{equation}\label{eq:did}
\Phi^{(a)} = \frac{1}{2}  ({}_a \phi  + \phi) = \phi - \frac{1}{2} \ln \sqrt{\frac{\det G_{a b}}{\det {}_a G^{a b}}} \, ,
\end{equation}
invariant under duality transformation along $x^a$ directions. It often occurs in the literature as "invariant dilaton" (see for example Ref.\cite{Hull2}).
The "doubled dilaton" $\Phi^{(a)}$  is good solution for the set of theories \cite{Hull,Hull2} where
for any  $d$  coordinates $x^a$  along which we perform T-duality it corresponds  the  actions
$S_d \,\, ,(d=1,2, \cdots, D)$.  These  actions are invariant under corresponding T-dualities.

We have one action which describes all  T-dual transformations and
we would like to have dilaton invariant under all possible T-duality transformations.
So, we will offer new expression for "doubled dilaton"
\begin{equation}\label{eq:ndo}
"\Phi" = \frac{1}{2^D} \sum_{a \in  A}   {}_a  \phi  = \phi - \ln \sqrt{\det G_{\mu \nu}} +   \frac{1}{2^D} \sqrt{ \prod_{a \in  A}  \det {}_a G_{\mu \nu}}  \, ,
\end{equation}
where index $A$ means summation (multiplication) of all possible $2^D$ T-dualities. It includes initial dilaton $\phi$ all $D$ T-dual dilatons along one of $D$ dimensions, all $D (D-1) \over 2$ T-dual dilatons along arbitrary two coordinates and so on, ending with T-dual dilaton along all dimensions. Using (\ref{eq:invd}) it is easy to show that above expression is duality invariant under all possible T-dualizations.
Because the last term is manifestly  duality invariant, we can take the first two terms as  invariant dilaton
\begin{equation}\label{eq:nd}
\Phi =  \phi - \ln \sqrt{\det G_{\mu \nu}}     \, .
\end{equation}
With the help of (\ref{eq:dsh}) we can check that it is duality invariant and we will take this expression as dilaton of the double space. In new notation the expression (\ref{eq:dsh}) takes the simple form
${}_a \Phi = \Phi$.

As before, with star we denote T-dualization along all coordinates.
Using (\ref{eq:gm}) and (\ref{eq:gmd}) we can  expressed generalized metric  symmetrically in term of initial metric and Kalb-Ramond  fields and their totally T-dual background fields
\begin{equation}\label{eq:gmid}
{\cal{H}}_{MN} (x,V) = \left (
\begin{array}{cc}
({}^{\diamond \star} G^{-1})_{\mu \nu}(V)   &  2 ({}^{\diamond \star} G^{-1})^{\mu \rho} (V) \, {}^{\diamond \star} B_{\rho \nu} (V)    \\
2 (G^{-1})^{\mu \rho} \, {}^{\diamond} B_{\rho \nu}(x) & (G^{-1})^{\mu \nu}
\end{array}\right )\, .
\end{equation}

Let us do a similar thing with dilaton field. Because we have ${}^\star \Phi = {}^\star \phi - \ln \sqrt{\det {}^\star G^{\mu \nu}}$,  we can express (\ref{eq:nd}) in symmetric form
in term of dilaton from the initial theory $\phi$ and dilaton from its totally T-dual theory ${}^\star \phi$
\begin{equation}\label{eq:nds}
\Phi = \frac{1}{2} (\Phi + {}^\star  \Phi)  =  \frac{1}{2} (\phi + {}^\star  \phi  ) - \frac{1}{2} \ln \sqrt{\det G_{\mu \nu}  \det {}^\star G^{\mu \nu}}     \, .
\end{equation}
This is useful relation for the path integral measure, because we can reexpress it as
\begin{equation}\label{eq:ndsm}
e^{ -2 \Phi} =   e^{- (\phi + {}^\star  \phi) } \sqrt{\det G_{\mu \nu}  \det {}^\star G^{\mu \nu}}     \, ,
\end{equation}
so that $e^{ -2 \Phi} d x^\mu d y_\mu$ is double space integration  measure, as well as in the Double field theory.


\cleq

\section{Example: Three torus}

In this section we will take the example of $3-$torus with $H$ flux background, considering  in of the Refs.\cite{ALLP,ALLP1,HLZ,BL,DNS2}. Using the method described above, we will compute backgrounds for all T-dual
theories, both geometric and non-geometric.

\subsection{Background of the initial theory}

Let us first fix the initial theory.
The coordinates of the $D=3$ dimensional torus will be denoted by $x^{1},x^{2},x^{3}$. In our particular example, nontrivial components of the  background are
\begin{equation}\label{eq:nasapolja}
G_{\mu\nu}=\delta_{\mu\nu}\, ,\quad B_{12}=-\frac{1}{2}Hx^3  \, ,
\end{equation}
or explicitly
\begin{eqnarray}
G_{\mu\nu}=\left(
\begin{array}{ccc}
1 & 0 & 0\\
0 & 1 & 0\\
0 & 0 & 1
\end{array}
\right)  \, , \qquad
B_{\mu\nu} (x)=\left(
\begin{array}{ccc}
0 & -\frac{1}{2} Hx^3 & 0\\
\frac{1}{2} Hx^3 & 0 & 0\\
0 & 0 & 0
\end{array}
\right) \, .
\end{eqnarray}
Here, the background fields depend only on $x^3$ and so in the whole section we will have $n=1$ and
\begin{equation}
P_r \to P_3= \left(\begin{array}{ccc}
0 & 0 & 0\\
0 & 0 & 0\\
0 & 0 & 1
\end{array} \right) \,.
\end{equation}

Note that the background fields considered in Refs.\cite{ALLP,ALLP1,HLZ,BL},
which will be marked by ${\cal G}$ and ${\cal B}$, are related with our notation as
\begin{equation}
{\cal B}_{\mu\nu}=-2B_{\mu\nu}\, ,\quad {\cal G}_{\mu\nu}=G_{\mu\nu}\, ,   \quad   (\mu,\nu=1,2,3) \,.
\end{equation}

It is easy to check that
\begin{eqnarray}
\Pi_{\pm \mu\nu} (x) \equiv B_{\mu\nu} (x) \pm \frac{1}{2} G _{\mu\nu}
=  \frac{1}{2} \left(
\begin{array}{ccc}
\pm 1 & - Hx^3  & 0\\
Hx^3  & \pm 1 & 0\\
0 & 0 & \pm 1
\end{array}
\right)   \, ,
\end{eqnarray}
and so,
\begin{equation}
G^E_{\mu\nu}=G_{\mu\nu} = \delta_{\mu\nu}\, ,\quad  \Theta_\pm^{\mu \nu} = - \frac{2}{\kappa} \,\,  \Pi_{\pm \mu\nu} \,  .
\end{equation}
Because $b_{\mu\nu}=0$ it follows that $\theta^{\mu\nu}_0 =0$.

For diamond background fields we have
\begin{eqnarray}
{}^{\diamond} B_{\mu\nu} (x)=\left(
\begin{array}{ccc}
0 & -\frac{3}{4} Hx^3 & 0\\
\frac{3}{4} Hx^3 & 0 & 0\\
0 & 0 & 0
\end{array}
\right) \, ,  \qquad \qquad
{}^{\diamond} \Pi_{\pm \mu\nu} (x) = \frac{1}{2} \left(
\begin{array}{ccc}
\pm 1 & - \frac{3}{2}  Hx^3  & 0\\
\frac{3}{2} Hx^3  & \pm 1 & 0\\
0 & 0 & \pm 1
\end{array}
\right)     \, .
\end{eqnarray}

\subsection{General form of T-duality transformations for Three torus}

In this example the T-duality chain for three dimensional space has the form
\begin{eqnarray}\label{eq:chain}
&&\Pi_{\pm \mu \nu}(x_{arg}),\,x^{\mu}
\mathrel{{\mathop{\mathop{\rightleftharpoons}^{\mathrm{T^1}}_{\mathrm{T_1}}}}}
\Pi_{1 \pm \mu \nu}(x_{1 arg}),\,x_1^{\mu}
\mathrel{{\mathop{\mathop{\rightleftharpoons}^{\mathrm{T^2}}_{\mathrm{T_2}}}}}
\Pi_{2 \pm \mu \nu}(x_{2 arg}),\,x_2^{\mu}
\mathrel{{\mathop{\mathop{\rightleftharpoons}^{\mathrm{T^3}}_{\mathrm{T_3}}}}}
\Pi_{3 \pm \mu \nu}(x_{3 arg}),\,x_3^{\mu}  \, ,
\end{eqnarray}
with coordinates
\begin{equation}
x^\mu= \{x^1, x^2, x^3 \}, \, \,\,  x_1^\mu= \{y_1, x^2, x^3 \}, \, \,\,  x_2^\mu= \{y_1, y_2, x^3 \},  \, \,\, x_3^\mu= \{y_1, y_2, y_3 \}  \, .
\end{equation}
The general forms of the arguments are
\begin{equation}
\begin{array}{cc}
x_{arg}^\mu= \{x^1, x^2, x^3 \}, \, \,\, \qquad  \qquad & x_{1 arg}^\mu= \{V^1(y_1), x^2, x^3 \}, \, \,\, \\ \\
 x_{2 arg}^\mu= \{V^1(y_1,y_2) , V^2(y_1,y_2), x^3 \}, \,\,\,  & x_{3 arg}^\mu= \{V^1(y_1,y_2,y_3), V^2(y_1,y_2,y_3), V^3(y_1,y_2,y_3) \} \, ,
 \end{array}
\end{equation}
where $V^a (y_a)$ will be calculated later as a  solutions of the zero order T-duality transformations.
Because in this example the arguments depend only on the third coordinate $(n=1)$ in the particular case, after projection with $P_r \to P_3$,   we have
\begin{equation}
x_{arg}^\mu= \{x^3 \}, \, \,\,  x_{1 arg}^\mu= \{x^3 \}, \, \,\,  x_{2 arg}^\mu= \{x^3 \}, \,\,\,    x_{3 arg}^\mu= \{ V^3 (y_1,y_2,y_3) \}  \, .
\end{equation}

In general, there are $ {D \choose 1} \to  {3 \choose 1} =3$ dual theories along one  coordinate, ${D \choose 2} \to  {3 \choose 2}=3$ dual theories along two coordinates and so on.
It is useful to represent  T-duality transformations between these theories  in the diagram
\begin{eqnarray}\label{eq:schain}
& {}_1 S (y_1, x^2, x^3)
\mathrel{{\mathop{\mathop{\longrightarrow}^{\mathrm{T^2}}}}}
 {}_{12} S (y_1, y_2, x^3)   &                                                 \nonumber \\
 \mathrel{{\mathop{\mathop{\nearrow}^{\mathrm{T^1}}}}}  &     \mathrel{{\mathop\nearrow^{\mathrm{T^1}}{\mathop{\searrow}^{\mathrm{T^3}}}}}           &   \mathrel{{\mathop{\mathop{\searrow}^{\mathrm{T^3}}}}}                  \nonumber \\
   S (x^1, x^2, x^3)
\mathrel{{\mathop{\mathop{\longrightarrow}^{\mathrm{T^2}}}}}
& {}_2 S (x^1, y_2, x^3)  \qquad  {}_{13} S (y_1, x^2, y_3)  &
\mathrel{{\mathop{\mathop{\longrightarrow}^{\mathrm{T^2}}}}}
{}_{123} S  (y_1, y_2, y_3,  V^3 )                                                       \nonumber \\
 \mathrel{{\mathop{\mathop{\searrow}^{\mathrm{T^3}}}}}        &  \mathrel{{\mathop\nearrow^{\mathrm{T^1}}{\mathop{\searrow}^{\mathrm{T^3}}}}}    &   \mathrel{{\mathop{\mathop{\nearrow}^{\mathrm{T^1}}}}}            \nonumber \\
& {}_3 S (x^1, x^2, y_3)
\mathrel{{\mathop{\mathop{\longrightarrow}^{\mathrm{T^2}}}}}
 {}_{23} S (x^1, y_2, y_3)    &\, .
\end{eqnarray}
It is possible to perform successive two and three T-dualizations as well as T-dualization in the directions opposite to the direction of the arrows. For simplicity we did not show it on the diagram.
The background fields for all theories of the above diagram will be collected in Subsection 6.8.
In the literature the theories $S, {}_1 S, {}_{12} S$ and ${}_{123} S$ are known as theories with $H,\, f \, , Q $ and $R$ fluxes respectively.

The explicit form of the double coordinate, the arguments of background fields  and the generalized metric is
\begin{equation}\label{eq:dc}
 Z^M =
  \left (
\begin{array}{c}
 x^\mu  \\
 x^\mu_3
\end{array}\right ) =
 \left (
\begin{array}{c}
 x^1  \\
 x^2   \\
 x^3    \\
 y_1  \\
 y_2   \\
 y_3
\end{array}\right )     \, , \qquad
 Z^M_{arg}
 =
 \left (
\begin{array}{c}
 x^\mu_{3 arg}    \\
  x^\mu_{3 arg}     \\
 x^\mu_{3 arg}    \\
 x^\mu_{arg}     \\
 x^\mu_{arg}       \\
  x^\mu_{arg}
  \end{array}\right )  =
 \left (
\begin{array}{c}
 V^3   \\
V^3     \\
V^3      \\
 x^3    \\
  x^3     \\
  x^3
\end{array}\right )  \, ,
\end{equation}

\begin{equation}\label{eq:gmp}
{\cal{H}}_{MN} ( Z^M_{arg})  = \left (
\begin{array}{cccccc}
1     &    0    &   0   &   0    &  \frac{3}{2}  H V^3      &    0            \\
0     &    1    &   0   &   -\frac{3}{2}  H V^3       &       &    0            \\
0     &    0    &   1   &   0    &  0    &    0            \\
0     &    -\frac{3}{2}  Hx^3      &   0   &   1    &       0            &    0            \\
 \frac{3}{2}  Hx^3       &    0    &   0   &   0    &        1         &    0            \\
0     &    0    &   0   &   0    &       0          &    1            \\
\end{array}\right )\, .
\end{equation}
Because in this example we have $n=1$, the $Z^M$ and $ Z^M_{arg}$ have the same dimension. Note that they transform with matrices ${\cal T}^a$ and $\breve {\cal T}^a$ respectively, which also have the same dimension,
but they are not equal.

\subsection{The expression for $V^3$  as solutions of zero order T-duality transformation}

In  expressions for $Z^M_{arg}$ (\ref{eq:dc}) and for ${\cal{H}}_{MN}$  (\ref{eq:gmp})  there is undetermined variable $V^3$   as
the arguments of background fields. It is solution of T-duality transformations at zero order.
In the flat space we have ${\cal{H}}_{MN} = 1$ and
\begin{equation}
\partial_\pm \,  Z^M = \pm \Omega \, Z^M \,  .
\end{equation}
Note that last relation does not depend on $d$. In general case, for different values of $d$, we should solve zero order T-duality transformations  independently.

There are three independent relations (in this example they have the same form)
\begin{equation}
\pm \partial_\pm \, y_\mu =  \partial_\pm \, x^\mu \, , \qquad  (\mu=1,2,3) .
\end{equation}
Because we should eliminate $y_i$ it is enough to solve  relations with index $a$. Its solution
\begin{equation}
 x^a = {\tilde y}_a \, ,  \qquad (a=1, \cdots d)
\end{equation}
where ${\tilde y}_a$ is defined in  (\ref{eq:dualnavar}),  we will proclaim as $V^a$.
In this example the arguments depend only on the third coordinate and we have
\begin{equation}
x_{arg}^\mu= \{x^3 \}, \, \,\,  x_{1 arg}^\mu= \{x^3 \}, \, \,\,  x_{2 arg}^\mu= \{x^3 \}, \,\,\,    x_{3 arg}^\mu= \{ V^3 ={\tilde y}_3 \}  \, .
\end{equation}

Now, we can replace $V^3$ with ${\tilde y}_3$ in the arguments of background fields  $Z^M_{arg}$  (\ref{eq:dc}) and in the generalized metric ${\cal{H}}_{MN}$ (\ref{eq:gm})  and obtain
\begin{equation}
 Z^M_{arg} =
 \left (
\begin{array}{c}
 {\tilde y}_3   \\
 {\tilde y}_3    \\
{\tilde y}_3     \\
 x^3    \\
  x^3     \\
  x^3
\end{array}\right )  \, , \qquad
{\cal{H}}_{MN} ( Z^M_{arg}) = \left (
\begin{array}{cccccc}
1     &    0    &   0   &   0    &  \frac{3}{2}  H {\tilde y}_3     &    0            \\
0     &    1    &   0   &   -\frac{3}{2}  H {\tilde y}_3      &       &    0            \\
0     &    0    &   1   &   0    &  0    &    0            \\
0     &    -\frac{3}{2}  Hx^3      &   0   &   1    &       0            &    0            \\
 \frac{3}{2}  Hx^3       &    0    &   0   &   0    &        1         &    0            \\
0     &    0    &   0   &   0    &       0          &    1            \\
\end{array}\right )\, .
\end{equation}

\subsection{$\hat {\cal{T}}^1$ :  H\,flux $\rightarrow$ f\, flux   (d=1)}

Let us first consider the case $d=1$, where the indices take the following values $a,b\in\{1\}$,  $i,j\in\{2,3\}$  and consequently $P_a \to P_1= \left (
\begin{array}{ccc}
1    &    0    &   0         \\
0     &    0    &   0         \\
0     &    0    &   0
\end{array}\right )  $.
It corresponds  to the T-dualization  along direction $x^{1}$,
$$ {\it \,\hat {\cal{T}}^1: torus\, with\, H\,flux \rightarrow torus\, with\, f\,flux (twisted\, torus)}\, , \qquad S_H [x^1, x^2, x^3]\rightarrow S_f [y_{1},x^{2},x^{3}] \, .$$

To perform such T-dualization  we should exchange $x^1$ with $y_1$ with matrix
\begin{equation}\label{eq:tau1}
({\cal{T}}^1)_{M}{}^N = \left (
\begin{array}{cccccc}
0     &    0    &   0   &   1    &  0     &    0            \\
0     &    1    &   0   &   0      &   0    &    0            \\
0     &    0    &   1   &   0    &  0    &    0            \\
1     &    0      &   0   &   0    &       0            &    0            \\
0     &    0    &   0   &   0    &        1         &    0            \\
0     &    0    &   0   &   0    &       0          &    1            \\
\end{array}\right ) = 1_2 \otimes (1-P_1) + \Omega_2 \otimes  P_1   \, ,
\end{equation}
and obtain instead of (\ref{eq:dualh1})
\begin{equation}\label{eq:tht1y}
{\cal{T}}^1   {\cal{H}}  {\cal{T}}^1 (Z_{{\cal{T}}^1   {\cal{H}}  {\cal{T}}^1}) =
\left (
\begin{array}{cccccc}
1     &    -\frac{3}{2}  Hx^3    &   0   &   0    &  0    &    0            \\
-\frac{3}{2}  H {\tilde y}_3     &    1    &   0   &   0      &       &    0            \\
0     &    0    &   1   &   0    &  0    &    0            \\
0     &    0    &   0   &   1    &      \frac{3}{2}  H {\tilde y}_3        &    0            \\
0      &    0    &   0   &   \frac{3}{2}  Hx^3    &        1         &    0            \\
0     &    0    &   0   &   0    &       0          &    1            \\
\end{array}\right )\, .
\end{equation}

Let us note that  after multiplication with ${\cal T}^1$    in the particular case (dependence only on the third coordinate) we have
\begin{equation}
Z_{\,  {\cal T}^1 {\cal{H}} \,\,  {\cal T}^1}^M =
\left (
\begin{array}{c}
 x^3    \\
{\tilde y}_3    \\
 /   \\
{\tilde y}_3  \\
 x^3   \\
 /
\end{array}\right )   \, .
\end{equation}
In this example we have only variables with index $3$ and in case of T-dualization along  $x^1$ the index $i$ takes the value $3$. So,
in order to obtain ${}_1 Z_{arg}$ we should change only variables with index $i$ which
according to the rule  (\ref{eq:cof})  appear in the rows $a=1$ and $a+D =4$. It produces
\begin{equation}
Z_{\, {\cal T}^1 {\cal{H}} \,\,  {\cal T}^1 }^M =
\left (
\begin{array}{c}
 x^3    \\
{\tilde y}_3    \\
 /  \\
{\tilde y}_3  \\
 x^3  \\
/
\end{array}\right )     \to
\left (
\begin{array}{c}
 {\tilde y}_3     \\
 {\tilde y}_3     \\
 /    \\
 x^3     \\
  x^3  \\
  /
\end{array}\right )  =
{}_1 Z_{arg}   = \breve{\cal S}^1  Z_{\, {\cal T}^1 {\cal{H}} \,\,  {\cal T}^1 }^M                   \, ,
\end{equation}
where in this case $\breve{\cal S}^1= {\cal T}^1$. So, instead of (\ref{eq:dualh2})  we obtain the generalized metric
\begin{equation}\label{eq:gm1}
 {}_1 {\cal{H}}_{MN} ({}_1 Z_{arg}) =
\left (
\begin{array}{cccccc}
1     &    -\frac{3}{2}  H {\tilde y}_3     &   0   &   0    &  0    &    0            \\
-\frac{3}{2}  H {\tilde y}_3     &    1    &   0   &   0      &       &    0            \\
0     &    0    &   1   &   0    &  0    &    0            \\
0     &    0    &   0   &   1    &      \frac{3}{2}  H x^3    &    0            \\
0      &    0    &   0   &   \frac{3}{2}  H x^3    &        1         &    0            \\
0     &    0    &   0   &   0    &       0          &    1            \\
\end{array}\right )\, .
\end{equation}

The expression for ${}_1 Z_{arg}$ we can also obtain directly multiplying $Z_{arg}$ with $\breve{{\cal T}}^1$.
Since, in our case $P_a P_r  \to P_1 P_3 =0$ from (\ref{eq:tsr})  we have $\breve{{\cal T}}^1 =1$ and ${}_1 Z_{arg}= Z_{arg}$.

According to (\ref{eq:hd}) the generalized metric  has a form (note that $i=3$)
\begin{equation}\label{eq:hd1}
{}_1 {\cal{H}}_{MN} ({}_1 Z_{arg}) = \left (
\begin{array}{cc}
{}_1  {}^{\diamond} G_E^{\mu \nu} (V^3={\tilde y}_3) &  -2  ({}_1  {}^{\diamond} B \,{}_1 {}^{\diamond} G^{-1})^\mu{}_\nu  (V^3={\tilde y}_3) \\
2 ({}_1 {}^{\diamond} G^{-1}\, {}_1  {}^{\diamond} B)_\mu {}^\nu (x^3)  &    ({}_1 {}^{\diamond} G^{-1})_{\mu \nu} (x^3)
\end{array}\right )\, .
\end{equation}

From lower $D=3$ rows we have
\begin{equation}
({}_1 {}^{\diamond} G^{-1})_{\mu \nu} (x^3) =
\left (
\begin{array}{ccc}
 1    &      \frac{3}{2}  H x^3    &    0            \\
\frac{3}{2}  H x^3    &        1         &    0            \\
 0    &       0          &    1            \\
\end{array}\right )\, ,  \qquad ({}_1 {}^{\diamond} G^{-1}\, {}_1  {}^{\diamond} B)_\mu {}^\nu =0 \, ,
\end{equation}
and consequently,
\begin{equation}
{}_1 {}^{\diamond} G^{\mu \nu} (x^3) =
\left (
\begin{array}{ccc}
 1    &      -\frac{3}{2}  H x^3    &    0            \\
-\frac{3}{2}  H x^3    &        1         &    0            \\
 0    &       0          &    1            \\
\end{array}\right )     \, ,        \qquad  {}_1  {}^{\diamond} B^{\mu \nu}  =0  \, .
\end{equation}
The upper $D=3$ rows produce the same result. So, the results without diamond
\begin{equation}\label{eq:GBf}
G^{\mu \nu}_f (x^3) \equiv
{}_1  G^{\mu \nu} (x^3) =
\left (
\begin{array}{ccc}
 1    &      -  H x^3    &    0            \\
- H x^3    &        1         &    0            \\
 0    &       0          &    1            \\
\end{array}\right )     \, ,        \qquad   B^{\mu \nu}_f  \equiv  {}_1  B^{\mu \nu}  =0  \, ,
\end{equation}
completely coincide with Refs.\cite{ALLP,ALLP1,HLZ,BL,DNS2}.

\subsection{$\hat {\cal{T}}^2$ :  f \, flux $\rightarrow$ Q \, flux (d=1)}

Let us consider the next $d=1$ case, where the indices take the following values $a,b\in\{2\}$, $i,j\in\{1,3\}$ and consequently $P_a \to P_2= \left (
\begin{array}{ccc}
0    &    0    &   0         \\
0     &    1    &   0         \\
0     &    0    &   0
\end{array}\right )  $.
It corresponds  to the T-dualization  along direction $x^{2}$,
$$ {\it \, \hat {\cal{T}}^2: torus\, with\, f\,flux \rightarrow torus\, with\, Q\,flux }\, , \qquad S_f [y_{1},x^{2},x^{3}] \rightarrow S_Q [y_{1}, y_{2},x^{3}]\, .$$

To perform such T-dualization  we exchange $x^2$ with $y_2$ with the matrix
\begin{equation}\label{eq:tau2}
({\cal{T}}^{2})_{M}{}^N = \left (
\begin{array}{cccccc}
1    &    0      &   0   &   0    &   0     &    0            \\
0     &    0     &   0   &   0    &   1     &    0            \\
0     &    0     &   1   &   0    &   0     &    0            \\
0     &    0     &   0   &   1    &   0     &    0            \\
0     &   1      &   0   &   0    &   0     &    0            \\
0     &    0     &   0   &   0    &   0     &    1            \\
\end{array}\right ) = 1_2 \otimes (1-P_2) + \Omega_2 \otimes  P_2     \,  ,
\end{equation}
and obtain
\begin{equation}\label{eq:tht2y}
{\cal{T}}^2 \, {}_1 {\cal{H}} \, {\cal{T}}^2 (Z_{{\cal{T}}^2 \, {}_1 {\cal{H}} \, {\cal{T}}^2} ) =
\left (
\begin{array}{cccccc}
1                              &    0                &   0   &   0              &  -\frac{3}{2}  H {\tilde y}_3    &    0            \\
0                             &   1                &   0   &   \frac{3}{2}  Hx^3  &    0                            &    0            \\
0                             &    0                 &   1   &   0               &  0                              &    0            \\
0                            &   \frac{3}{2}  Hx^3   &   0   &   1               & 0                         &    0            \\
-\frac{3}{2}  H {\tilde y}_3  &    0                  &   0   &   0                &        1                      &    0            \\
0                            &    0                  &   0   &   0                 &       0                     &    1            \\
\end{array}\right )\, .
\end{equation}

According to (\ref{eq:cof}) we should again change only variables with index $i$  at rows $a=2$ and $a+D=5$
\begin{equation}
Z_{\, {\cal T}^2 {}_1 {\cal{H}} \,\,  {\cal T}^2 }^M =
\left (
\begin{array}{c}
{\tilde y}_3     \\
x^3    \\
/    \\
 x^3  \\
{\tilde y}_3  \\
/
\end{array}\right )     \to
\left (
\begin{array}{c}
 {\tilde y}_3     \\
 {\tilde y}_3     \\
 /     \\
 x^3     \\
  x^3  \\
 /
\end{array}\right )  =
{}_{12} Z_{arg}  = \breve{{\cal S}}^2 Z_{\, {\cal T}^2 {}_1 {\cal{H}} \,\,  {\cal T}^2 }^M      \, ,
\end{equation}
(where again $\breve{\cal S}^2= {\cal T}^2$)  so that the generalized metric takes the final form
\begin{equation}\label{eq:gm2}
 {}_{12} {\cal{H}}_{MN} ({}_{12} Z_{arg}) =
 \left (
\begin{array}{cccccc}
1                              &    0                &   0   &   0              &  -\frac{3}{2}  H {\tilde y}_3    &    0            \\
0                             &   1                &   0   &   \frac{3}{2}  H {\tilde y}_3    &    0                            &    0            \\
0                             &    0                 &   1   &   0               &  0                              &    0            \\
0                            &   \frac{3}{2}  H x^3   &   0   &   1               & 0                         &    0            \\
-\frac{3}{2}  H x^3  &    0                  &   0   &   0                &        1                      &    0            \\
0                            &    0                  &   0   &   0                 &       0                     &    1            \\
\end{array}\right )\, .
\end{equation}
Again we have $\breve{{\cal T}}^2 =1$ and ${}_{12} Z_{arg}={}_1 Z_{arg}$. According to (\ref{eq:hd}) the generalized metric has a form
\begin{equation}\label{eq:hd2}
{}_{12} {\cal{H}}_{MN} ({}_{12} Z_{arg})= \left (
\begin{array}{cc}
{}_{12}  {}^{\diamond} G_E^{\mu \nu} (V^3={\tilde y}_3) &  -2  ({}_{12}  {}^{\diamond} B \,{}_{12} {}^{\diamond} G^{-1})^\mu{}_\nu  (V^3={\tilde y}_3) \\
2 ({}_{12} {}^{\diamond} G^{-1}\, {}_{12}  {}^{\diamond} B)_\mu {}^\nu (x^3)  &    ({}_{12} {}^{\diamond} G^{-1})_{\mu \nu} (x^3)
\end{array}\right )\, .
\end{equation}

From lower $D=3$ rows we have
\begin{equation}
({}_{12} {}^{\diamond} G^{-1})_{\mu \nu}  =
\left (
\begin{array}{ccc}
 1    &      0    &    0            \\
0    &        1         &    0            \\
 0    &       0          &    1            \\
\end{array}\right )\, ,  \qquad
({}_{12} {}^{\diamond} G^{-1}\, {}_{12}  {}^{\diamond} B)_\mu {}^\nu (x^3) =
\left (
\begin{array}{ccc}
 0   &      \frac{3}{4}  H x^3    &    0            \\
-\frac{3}{4}  H x^3    &        0         &    0            \\
 0    &       0     &       0
 \end{array}\right )\, ,
\end{equation}
and consequently,
\begin{equation}
{}_{12} {}^{\diamond} G^{\mu \nu}  =
\left (
\begin{array}{ccc}
 1    &      0   &    0            \\
0  &        1         &    0            \\
 0    &       0          &    1            \\
\end{array}\right )     \, ,        \qquad
{}_{12}  {}^{\diamond} B^{\mu \nu} (x^3) =
 \left (
\begin{array}{ccc}
 0   &      \frac{3}{4}  H x^3    &    0            \\
-\frac{3}{4}  H x^3    &        0         &    0            \\
 0    &       0     &       0
 \end{array}\right ) \, .
\end{equation}
The upper $D=3$ rows produce the same result. So, the results without diamond
\begin{equation}\label{eq:GBQ}
G^{\mu \nu}_Q  \equiv
{}_{12}  G^{\mu \nu}  =
\left (
\begin{array}{ccc}
 1    &      0    &    0            \\
0 &        1         &    0            \\
 0    &       0          &    1            \\
\end{array}\right )     \, ,        \qquad
B^{\mu \nu}_Q (x^3) \equiv  {}_{12}  B^{\mu \nu} (x^3) =
 \left (
\begin{array}{ccc}
 0   &      \frac{1}{2}  H x^3    &    0            \\
-\frac{1}{2}  H x^3    &        0         &    0            \\
 0    &       0     &       0
 \end{array}\right )  \, ,
\end{equation}
completely coincide with Refs.\cite{ALLP,ALLP1,HLZ,BL,DNS2}.

\subsection{$\hat {\cal{T}}^{12} $ :  H \, flux $\rightarrow$ Q \, flux  (d=2)}

Let us reproduce the result of the previous two subsections with one transformation.
We will consider the case $d=2$, where the indices take the following values $a,b\in\{1, 2\}$ and $i,j\in\{3\}$ and consequently $P_a \to P_{12}= \left (
\begin{array}{ccc}
1    &    0    &   0         \\
0     &    1    &   0         \\
0     &    0    &   0
\end{array}\right )  $.
It corresponds  to the T-dualization  along direction $x^{1}$ and  $x^{2}$
$$ {\it \,\hat {\cal{T}}^{12}: torus\, with\, H\,flux \rightarrow torus\, with\, Q\,flux \, , \qquad S_H [x^{1},x^{2},x^{3}] \rightarrow S_Q [y_{1}, y_{2},x^{3}]} \, .$$

To perform such T-dualization  we should exchange $x^1$ with $y_1$ and $x^2$ with $y_2$ with the matrix
\begin{equation}\label{eq:tau12}
{\cal{T}}^{12} = {\cal{T}}^{1} {\cal{T}}^{2}  = \left (
\begin{array}{cccccc}
0     &    0    &   0   &   1    &  0     &    0            \\
0     &    0    &   0   &   0      &   1    &    0            \\
0     &    0    &   1   &   0    &  0    &    0            \\
1     &    0      &   0   &   0    &       0            &    0            \\
0     &   1    &   0   &   0    &       0       &    0            \\
0     &    0    &   0   &   0    &       0          &    1            \\
\end{array}\right )  = 1_2 \otimes (1-P_{12}) + \Omega_2 \otimes  P_{12}    \, ,
\end{equation}
and obtain
\begin{equation}\label{eq:tht12y}
{\cal{T}}^{12}  {\cal{H}}  {\cal{T}}^{12} (Z_{{\cal{T}}^{12}  {\cal{H}}  {\cal{T}}^{12}}) =
\left (
\begin{array}{cccccc}
1                              &    0                &   0   &   0              &  -\frac{3}{2}  H x^3     &    0            \\
0                             &   1                &   0   &   \frac{3}{2}  H x^3  &    0                            &    0            \\
0                             &    0                 &   1   &   0               &  0                              &    0            \\
0                            &   \frac{3}{2}  H {\tilde y}_3   &   0   &   1               & 0                         &    0            \\
-\frac{3}{2}  H {\tilde y}_3  &    0                  &   0   &   0                &        1                      &    0            \\
0                            &    0                  &   0   &   0                 &       0                     &    1            \\
\end{array}\right )\, .
\end{equation}

According to (\ref{eq:cof}) we should change only variables with index $i=3$  at rows  $a=1,2$ and $a+D = 4,5$
\begin{equation}
Z_{\, {\cal T}^{12} {\cal{H}} \,\,  {\cal T}^{12} }^M =
\left (
\begin{array}{c}
x^3      \\
x^3    \\
/   \\
{\tilde y}_3  \\
{\tilde y}_3  \\
/
\end{array}\right )     \to
\left (
\begin{array}{c}
 {\tilde y}_3     \\
 {\tilde y}_3     \\
/     \\
 x^3     \\
  x^3  \\
 /
\end{array}\right )  =
{}_{12} Z_{arg}  = \breve{\cal S}^{12} Z_{\, {\cal T}^{12} {\cal{H}} \,\,  {\cal T}^{12} }^M     \, ,
\end{equation}
(where  $\breve{\cal S}^{12}= {\cal T}^{12}$)  so that the generalized metric ${}_{12} {\cal{H}}_{MN} ({}_{12} Z_{arg})$ takes the same final form as in (\ref{eq:gm2}).  Again we have $\breve{{\cal T}}^{12} =1$ and ${}_{12} Z_{arg}= Z_{arg}$.
Consequently, the final result  (\ref{eq:GBQ}) is the same as in previous subsection.

\subsection{Non-geometric theory with R-flux in three ways  \\
 $\hat {\cal{T}}^3$: Q \, flux $\rightarrow$ R \, flux (d=1), $\hat {\cal{T}}^{23}$: f \, flux $\rightarrow$ R \, flux (d=2)  and \\
  $\hat {\cal{T}}^{123}$: H \, flux $\rightarrow$ R \, flux (d=3)}

We are going to obtain the background fields for theory with $R$ flux in three different ways. First, in the case (d=1) for  $a,b\in \{3 \}$,  $i,j\in\{1,2\}$ where  $P_a \to P_{3}= \left (
\begin{array}{ccc}
0    &    0    &   0         \\
0     &    0    &   0         \\
0     &    0    &   1
\end{array}\right )  $.
Applying   the matrix
\begin{equation}\label{eq:tau3}
({\cal{T}}^{3})_{M}{}^N = \left (
\begin{array}{cccccc}
1     &    0    &   0   &   0    &  0     &    0            \\
0     &    1   &    0   &   0    &   0    &    0            \\
0     &    0    &   0   &   0    &  0     &    1            \\
0     &    0    &   0   &   1    &  0     &    0            \\
0     &   0     &   0   &   0    &   1    &    0            \\
0     &    0    &   1   &   0    &   0   &    0            \\
\end{array}\right )  = 1_2 \otimes (1-P_{3}) + \Omega_2 \otimes  P_{3}  \, ,
\end{equation}
on the  $ {}_{12} {\cal{H}}_{MN} ({}_{12} Z_{arg})$, which has been obtained in (\ref{eq:gm2}), we find
\begin{equation}\label{eq:tht3y}
{\cal{T}}^{3} \,  {}_{12} {\cal{H}} \, {\cal{T}}^{3} (Z_{{\cal{T}}^{3} \,  {}_{12} {\cal{H}} \, {\cal{T}}^{3}}) =
\left (
\begin{array}{cccccc}
1                              &    0                &   0   &   0              &  -\frac{3}{2}  H {\tilde y}_3     &    0            \\
0                             &   1                &   0   &   \frac{3}{2}  H {\tilde y}_3 &    0                            &    0            \\
0                             &    0                 &   1   &   0               &  0                              &    0            \\
0                            &   \frac{3}{2}  H x^3   &   0   &   1               & 0                         &    0            \\
-\frac{3}{2}  H x^3    &    0                  &   0   &   0                &        1                      &    0            \\
0                            &    0                  &   0   &   0                 &       0                     &    1            \\
\end{array}\right )\, .
\end{equation}
Now, we should change the variables with index $a$ (because now $a=3$). According to (\ref{eq:cof}) they appear in the rows $i=1,2$ and $i+D=4,5$
 and we find
\begin{equation}\label{eq:gm3}
{}_{123} {\cal{H}}_{MN} ({}_{123} Z_{arg})  =
\left (
\begin{array}{cccccc}
1                              &    0                &   0   &   0              &  -\frac{3}{2}  H  x^3     &    0            \\
0                             &   1                &   0   &   \frac{3}{2}  H  x^3  &    0                            &    0            \\
0                             &    0                 &   1   &   0               &  0                              &    0            \\
0                            &   \frac{3}{2}  H {\tilde y}_3   &   0   &   1               & 0                         &    0            \\
-\frac{3}{2}  H {\tilde y}_3     &    0                  &   0   &   0                &        1                      &    0            \\
0                            &    0                  &   0   &   0                 &       0                     &    1            \\
\end{array}\right )\, .
\end{equation}

In all previous cases (for $a \ne 3$) we had $\breve{{\cal T}}^a=1$. In all cases of this subsection we have $P_{a=3} P_{r=3} = P_{a=2,3} P_{r=3} = P_{a=1,2,3} P_{r=3} = P_3$ and consequently
$\breve{{\cal T}}^3 = \breve{{\cal T}}^{2,3}= \breve{{\cal T}}^{1,2,3}= \Omega_2 \otimes 1_3 \otimes P_3$. It means that we should just exchange variables
from upper $D=3$ rows with that from lower $D=3$ rows, in particular $x^3$ with ${\tilde y}_3$. To the same conclusion we can come using (\ref{eq:t2t1}) and the fact that
$\breve{\cal S}^{3}= \bar {\cal T}^3$.

The same result we can obtain applying ${\cal{T}}^{23}= {\cal{T}}^{2} {\cal{T}}^{3}$ on  ${}_1 {\cal{H}}_{MN} ({}_1 Z_{arg}) $, which has been obtained in  (\ref{eq:gm1}),  in the case (d=2) for  $a,b\in \{2,3 \}$ and $i,j\in\{1 \}$, as well as
applying
\begin{equation}\label{eq:tau123}
({\cal{T}}^{1 2 3})_{M}{}^N = ({\cal{T}}^{1}  {\cal{T}}^{2} {\cal{T}}^{3})_{M}{}^N  = \left (
\begin{array}{cccccc}
0     &    0    &   0   &   1    &  0     &    0            \\
0     &    0    &   0   &   0      &   1    &    0            \\
0     &    0    &   0   &   0    &  0    &    1            \\
1     &    0      &   0   &   0    &       0            &    0            \\
0     &   1    &   0   &   0    &       0       &    0            \\
0     &    0    &   1   &   0    &       0          &    0            \\
\end{array}\right ) = \left (
\begin{array}{cc}
0     &    1_3           \\
1_3     &    0              \\
\end{array}\right )
\, ,
\end{equation}
on the starting generalized metric ${\cal{H}}_{MN}$ (\ref{eq:gm}).

The obtaining generalized metric should has a form (\ref{eq:hd})
\begin{equation}\label{eq:hd3}
{}_{123} {\cal{H}}_{MN} ({}_{123} Z_{arg})= \left (
\begin{array}{cc}
{}_{123}  {}^{\diamond} G_E^{\mu \nu} (x^3)  &  -2  ({}_{123}  {}^{\diamond} B \,{}_{123} {}^{\diamond} G^{-1})^\mu{}_\nu  (x^3)  \\
2 ({}_{123} {}^{\diamond} G^{-1}\, {}_{123}  {}^{\diamond} B)_\mu {}^\nu (V^3={\tilde y}_3)   &    ({}_{123} {}^{\diamond} G^{-1})_{\mu \nu} (V^3={\tilde y}_3)
\end{array}\right )\, .
\end{equation}
Note that in general, according to  (\ref{eq:hd}), all background fields in upper $D$ rows depend on $x^a$ and $V^i$ while in
lower $D$ rows depend on $V^a$ and $x^i$. Up to now we had that index $i$ took value $3$, but now first time the index $a$ takes value $3$.
So, the background fields in the upper $3$ rows depend on $x^3$ and in the lower $3$ rows on $V^3={\tilde y}_3$,
the opposite of the previous cases.
Consequently, the background fields for theory with $R$ flux, which take the position of lower $3$ rows, will depend on $V^3={\tilde y}_3$.
This is something that we could expect, because we realized  T-duality along $x^3$ direction on which background fields depend.
The result of this subsection can not be obtained with standard Buscher approach, but the generalized one of Refs.\cite{DS1,DNS2} must be applied.

With the same reason as in previous sections we have
\begin{equation}\label{eq:GBR}
G^{\mu \nu}_R  \equiv
{}_{123}  G^{\mu \nu}  =
\left (
\begin{array}{ccc}
 1    &      0    &    0            \\
0 &        1         &    0            \\
 0    &       0          &    1            \\
\end{array}\right )     \, ,        \qquad
B^{\mu \nu}_R ({\tilde y}_3) \equiv  {}_{123}  B^{\mu \nu} ({\tilde y}_3) =
 \left (
\begin{array}{ccc}
 0   &      \frac{1}{2}  H {\tilde y}_3    &    0            \\
-\frac{1}{2}  H {\tilde y}_3     &        0         &    0            \\
 0    &       0     &       0
 \end{array}\right )  \, ,
\end{equation}
which completely coincide with Ref.\cite{DNS2}. Unlike previous cases here the background is non-geometric and it  depends on ${\tilde y}_3$.

\subsection{Collection of all metrics and Kalb-Ramond fields for Three torus}

In a similar way as described above we can find all metrics and Kalb-Ramond fields for Three torus. Here  we will collect the final results for all theories in diagram (\ref{eq:schain})

\begin{eqnarray}\label{eq:d0}
d=0: \qquad
G_{\mu\nu}=\left(
\begin{array}{ccc}
1 & 0 & 0\\
0 & 1 & 0\\
0 & 0 & 1
\end{array}
\right)  \, , \qquad
B_{\mu\nu} (x)=\left(
\begin{array}{ccc}
0 & -\frac{1}{2} Hx^3 & 0\\
\frac{1}{2} Hx^3 & 0 & 0\\
0 & 0 & 0
\end{array}
\right) \, ,
\end{eqnarray}

\begin{eqnarray}\label{eq:d1}
 d=1: &     \nonumber \\
& {}_1  G^{\mu \nu} (x^3)  =
\left (
\begin{array}{ccc}
 1    &      -  H x^3    &    0            \\
- H x^3    &        1         &    0            \\
 0    &       0          &    1            \\
\end{array}\right )     \, ,          & {}_1  B^{\mu \nu}  =  0  \, ,                  \nonumber \\
& {}_2  G^{\mu \nu} (x^3)  =
\left (
\begin{array}{ccc}
 1    &        H x^3    &    0            \\
 H x^3    &        1         &    0            \\
 0    &       0          &    1            \\
\end{array}\right )     \, ,           & {}_2  B^{\mu \nu}  =  0  \, , \nonumber \\
& {}_{3}  G^{\mu \nu}  =
\left (
\begin{array}{ccc}
 1    &      0    &    0            \\
0 &        1         &    0            \\
 0    &       0          &    1            \\
\end{array}\right )     \, ,
& {}_{3}  B^{\mu \nu} ({\tilde y}_3)  =
 \left (
\begin{array}{ccc}
 0   &      -\frac{1}{2}  H {\tilde y}_3     &    0            \\
\frac{1}{2}  H {\tilde y}_3     &        0         &    0            \\
 0    &       0     &       0
 \end{array}\right )  \, ,
\end{eqnarray}

\begin{eqnarray}\label{eq:d2}
 d=2: &     \nonumber \\
& {}_{12}  G^{\mu \nu}  =
\left (
\begin{array}{ccc}
 1    &      0    &    0            \\
0 &        1         &    0            \\
 0    &       0          &    1            \\
\end{array}\right )     \, ,        \qquad
& {}_{12}  B^{\mu \nu} (x^3) =
 \left (
\begin{array}{ccc}
 0   &      \frac{1}{2}  H x^3    &    0            \\
-\frac{1}{2}  H x^3    &        0         &    0            \\
 0    &       0     &       0
 \end{array}\right )  \, ,                  \nonumber \\
&  {}_{23}  G^{\mu \nu} ({\tilde y}_3) =
\left (
\begin{array}{ccc}
 1    &        H {\tilde y}_3    &    0            \\
 H {\tilde y}_3    &        1         &    0            \\
 0    &       0          &    1            \\
\end{array}\right )     \, ,        \qquad   & {}_{23}  B^{\mu \nu}  =0  \, ,                 \nonumber \\
& {}_{13}  G^{\mu \nu} ({\tilde y}_3) =
\left (
\begin{array}{ccc}
 1    &       - H {\tilde y}_3    &    0            \\
 - H {\tilde y}_3    &        1         &    0            \\
 0    &       0          &    1            \\
\end{array}\right )     \, ,        \qquad  &  {}_{13}  B^{\mu \nu}  =0  \, ,
\end{eqnarray}

\begin{equation}\label{eq:d3}
d=3: \qquad
{}_{123}  G^{\mu \nu}  =
\left (
\begin{array}{ccc}
 1    &      0    &    0            \\
0 &        1         &    0            \\
 0    &       0          &    1            \\
\end{array}\right )     \, ,        \qquad
 {}_{123}  B^{\mu \nu} ({\tilde y}_3) =
 \left (
\begin{array}{ccc}
 0   &      \frac{1}{2}  H {\tilde y}_3    &    0            \\
-\frac{1}{2}  H {\tilde y}_3     &        0         &    0            \\
 0    &       0     &       0
 \end{array}\right )  \, .
\end{equation}

In the particular example, up to higher order of $H^2$ which we will neglected, we have $\det {}_a G_{\mu \nu}=1$ which produces  ${}_a \phi = \phi$. So,  dilaton field is invariant under T-duality transformations and doubled  dilaton  is equal to the initial dilaton, $\Phi= \phi$.

All theories obtained after dualization along $x^3$ coordinate (the coordinate on which  background depends) are nongeometric. Here they are ${}_3 S, {}_{13} S, {}_{23} S$ and ${}_{123} S$. The other theories
$S, {}_{1} S, {}_{2} S$ and ${}_{12} S$ are geometric.

\subsection{T-duality group $G_T(3)$}

Let us illustrate the T-duality group in the case of three torus. The three matrices ${\cal{T}}^1\, , {\cal{T}}^2$ and  ${\cal{T}}^3$, defined in the relations (\ref{eq:tau1}), (\ref{eq:tau2}) and (\ref{eq:tau3}),
are generators of the group $G_T(3)$. This group is subgroup of the $6-$permutational group. These matrices  can be represent in the following way
\begin{equation}\label{eq:pg3}
{\cal{T}}^1:   \left (
\begin{array}{cccccc}
1           &   2   & 3 & 4  &   5    & 6 \\
4      &  2   & 3 & 1       & 5    & 6
\end{array}\right )  \quad
{\cal{T}}^2:   \left (
\begin{array}{cccccc}
1           &   2   & 3 & 4  &   5    & 6 \\
1      &  5   & 3 & 4       & 2    & 6
\end{array}\right )  \quad
{\cal{T}}^3:   \left (
\begin{array}{cccccc}
1           &   2   & 3 & 4  &   5    & 6 \\
1      &  2   & 6 & 4       & 5    & 3
\end{array}\right )  \, ,
\end{equation}
or in the cyclic notation
\begin{equation}\label{eq:cpg}
{\cal{T}}^1: (1,4) \quad  {\cal{T}}^2:  (2,5) \quad  {\cal{T}}^2: (3, 6)   \, .
\end{equation}
The whole group has $8$ elements: $1, {\cal{T}}^1, {\cal{T}}^2, {\cal{T}}^3, {\cal{T}}^{12}, {\cal{T}}^{13}, {\cal{T}}^{23}$ and ${\cal{T}}^{123}$ and all  matrices  can be simply obtained by matrix multiplication of the  generators, for example ${\cal{T}}^{12}= {\cal{T}}^1 {\cal{T}}^2$.

The only independent "breve" matrices are $\breve{{\cal{T}}^1}= \breve{{\cal{T}}^2}=1_6$, and  $\breve{{\cal{T}}^3}= \Omega_2 \otimes 1_3 $. They form a group isomorphic to the group of two elements $1$ and $\Omega_2$.
In fact we have $\breve{{\cal{T}}^1}= \breve{{\cal{T}}^2}= \breve{{\cal{T}}}^{12}=1_6$ and $\breve{{\cal{T}}^3}= \breve{{\cal{T}}}^{13}=\breve{{\cal{T}}}^{23}=\breve{{\cal{T}}}^{123}=\Omega_2 \otimes 1_3$. They can be represent as
\begin{equation}\label{eq:bpg3}
{\breve{\cal T}}^1:   \left (
\begin{array}{cccccc}
1           &   2   & 3 & 4  &   5    & 6 \\
1      &  2   & 3 & 4       & 5    & 6
\end{array}\right )  \qquad
{\breve{\cal T}}^3:   \left (
\begin{array}{cccccc}
1           &   2   & 3 & 4  &   5    & 6 \\
4      &  5   & 6 & 1       & 2    &   3
\end{array}\right )   \, .
\end{equation}


\section{Conclusion}

In the paper \cite{SB} we offered simple formulation for T-duality transformations.
We introduced the extended $2D$ dimensional  space with the coordinates $Z^M= (x^\mu, y_\mu)$, which beside initial  $D$ dimensional space-time coordinates $x^\mu$ contains
the corresponding T-dual coordinates $y_\mu$.
We showed that in that double space the T-duality along some subset of coordinates $x^a$ $(a=0,1, \cdots ,d-1)$ and corresponding dual coordinates $y_a$ is equivalent to replacing their places.

Here we generalize this result to the case of weakly curved background where in addition to the extended coordinate we should also transform extended
argument of the background fields.
We define particular permutation of the coordinates, realized by operator   $\hat {\cal T}^a$, which  in the weakly curved background has two roles. First, with matrices ${\cal T}^a$  it exchanges the places of some subset of the coordinates $x^a$ and the corresponding dual coordinates $y_a$ along which we perform T-dualization. Second, in arguments of background fields $Z_{arg} = [V^\mu (y), x^\mu]$
it exchanges $x^a$ with its T-dual image $V^a(y)$, with matrices $\breve{{\cal T}}^a$. Matrices ${\cal T}^a$ and $\breve{{\cal T}}^a$ are homeomorphic and they are representation of the operator $\hat {\cal T}^a$.

We require that the obtained double space coordinates satisfy the same form of T-duality transformations as the initial one,
or in other words that this permutation is a global symmetry of the T-dual transformation.
We show that this permutation produce exactly the same T-dual background fields as
in the generalized Buscher approach of Ref.\cite{DNS2}.

In the flat space-time, this statement has been proved by direct calculations in Ref.\cite{SB}.
In the weakly curved background, thanks to the arguments of background fields,  we made nontrivial generalization.
But in that case we should solve the problem iteratively.
We start with T-duality transformations
(\ref{eq:tdualwc}), with  variables $Z^M$ and $Z_{arg}$. The
zero order does not depend on $Z_{arg}$,  because it appears with infinitesimal coefficient.
Elimination of $y_i$ from T-duality equations produce solution $x^a(x^i,y_a)$, eq.(\ref{eq:pard0s}), while elimination of $y_a$ produce solution $x^i(x^a, y_i)$, eq.(\ref{eq:pard1s}).
The solution  for $x^a$  we proclaim the  $V^a(x^i,y_a)$ and solution  for $x^i$  we proclaim the $V^i(x^a, y_i)$.
Thanks to the plus minus sign in front of T-duality relations (\ref{eq:tdualwc})
these solutions depend on both $y_a$ and its double ${\tilde y}_a$.
The zero order background field arguments  $Z_{arg}^{(0)}$ consists of $V^\mu$ and $x^\mu$, see (\ref{eq:dsvp}).
It is the arguments of the first order transformations, because it appears with infinitesimal coefficient. So,
we obtain the first order transformations, which according to (\ref{eq:tdual1}), produce equations of motion.

In the standard Buscher formulation T-duality transforms the initial theory to the equivalent one, T-dual theory. The double space formulation contains
both initial and T-dual theory and T-duality becomes the global symmetry transformation. With the help of (\ref{eq:dualgm})
it is,  easy to see that equations of motion (\ref{eq:tdual1}) are invariant under transformation
$Z^M \to Z^{\prime M} = ({\cal T}^a )^M{}_N Z^N\, , \quad Z_{arg} \to Z^{\prime }_{arg} = \breve{{\cal T}}^a  Z_{arg}$.

We have shown that in the case of weakly curved background, although the path integral is not Gaussian,  the T-duality transformation for the dilaton formally has the same form as in the case of the flat background,
where the path integral is Gaussian. As well as in the other approaches, the double dilaton $\Phi$ is duality invariant. In our approach it is invariant with respect to all T-duality transformations.
 We have expressed it in term of dilaton from the initial theory $\phi$ and dilaton from its totally T-dual theory ${}^\star \phi$. This is analogously to the generalized metric, which is expressed symmetrically
 in terms of metric and Kalb-Ramond  fields of  both initial and its totally T-dual theory.

Because both $Z^M$ and $Z_{arg}$ are transformed with the homeomorphic matrices  ${\cal T}^a$ and  $\breve{{\cal T}}^a$, the T-duality group with respect to the successive T-dualizations  is the same as in the case of flat background. It
is a subgroup of the $2D$ permutation group, which permute some $d$ of the first $D$ coordinates with corresponding $d$ of the last $D$ coordinates. In the cyclic form
it can be written as
\begin{equation}\label{eq:cpg1}
(1,D+1) (2,D+2) \cdots (d, D+d)   \, ,   \qquad  d \in  (0 ,D) \,  ,
\end{equation}
where  $d=0$ formally corresponds to the neutral element (no permutations) and $d=D$ corresponds to the case when T-dualization is performed along all coordinates.

It is well known that, in the case when background fields depend on all coordinates, all theories in the chain (\ref{eq:chain2}), except the initial one, are nongeometric. In the case of the weakly curved
background, the source of the nongeometry is  nonlocality of the arguments, which itself  is a line integral. It
depend on two variables, the Lagrangian multiplier $y_\mu$ and its double ${\tilde y}_\mu$. So, because our approach unify all nodes of the chain (\ref{eq:chain2}) it unify geometric with non-geometric theories.
It is also clearly explains that T-duality is unphysical, because it is equivalent to the permutation
of some coordinates.

Our approach is essentially different from the standard one of the double field theory, where all background fields depend on the same variable $Z^M$.
Here, it formally depends on two kind of double space coordinates $Z^M$ and $Z_{arg}$. The argument  $Z_{arg}$ appears with infinitesimal coefficient and it is solution for $Z^M$ in one order smaller approximation.
So, it essentially depend only on variable $Z^M$, but we should solve the problem  iteratively.

\appendix
\cleq

\section{Block-wise expressions for background fields}\label{sec:dodatak}

In order to simplify notation and to write expressions without indices (as matrix multiplication) we will introduce notations for
component fields.


\subsection{Block-wise expressions for flat background fields}

For the metric tensor and the Kalb-Ramond background fields we define
\begin{equation}\label{eq:G}
G_{\mu \nu} = \left (
\begin{array}{cc}
{\tilde G}_{ab}    &    G_{aj}       \\
 G_{ib}            &   {\bar G}_{ij}
\end{array}\right )   \equiv
 \left (
\begin{array}{cc}
{\tilde G}    &    G^T       \\
 G            &   {\bar G}
\end{array}\right )   \, ,
\end{equation}
and
\begin{equation}
b_{\mu \nu} = \left (
\begin{array}{cc}
{\tilde b}_{ab}    &    b_{aj}       \\
 b_{ib}            &   {\bar b}_{ij}
\end{array}\right )   \equiv
 \left (
\begin{array}{cc}
{\tilde b}    &    -b^T       \\
 b            &   {\bar b}
\end{array}\right )   \, .
\end{equation}
We also define
\begin{equation}\label{eq:gnmj}
(G^{-1})^{\mu \nu} = \left (
\begin{array}{cc}
{\tilde \gamma}^{ab}    &    \gamma^{aj}       \\
 \gamma^{ib}            &   {\bar \gamma}^{ij}
\end{array}\right )   \equiv
 \left (
\begin{array}{cc}
{\tilde \gamma}    &    \gamma^T       \\
 \gamma          &   {\bar \gamma}
\end{array}\right )   \, ,
\end{equation}
and the effective metric
\begin{equation}\label{eq:gdef}
g_{\mu \nu} = G_{\mu \nu} -4 b_{\mu\rho} (G^{-1})^{\rho \sigma} b_{\sigma \nu}
= \left (
\begin{array}{cc}
{\tilde g}_{ab}    &    g_{aj}       \\
 g_{ib}            &   {\bar g}_{ij}
\end{array}\right )   \equiv
 \left (
\begin{array}{cc}
{\tilde g}    &    g^T       \\
 g            &   {\bar g}
\end{array}\right )   \, .
\end{equation}

Note that because $G^{\mu \nu}$ is inverse of $G_{\mu \nu}$  we have
\begin{eqnarray}\label{eq:cgama}
& \gamma = - {\bar G}^{-1} G {\tilde \gamma} = - {\bar \gamma} G {\tilde G}^{-1} \, , \qquad
& \gamma^T = - {\tilde G}^{-1} G^T {\bar \gamma} = - {\tilde \gamma} G^T {\bar G}^{-1} \, , \nonumber \\
& {\tilde \gamma} = ({\tilde G}- G^T {\bar G}^{-1} G)^{-1} \,  ,
& {\bar \gamma} = ({\bar G}- G {\tilde G}^{-1} G^T)^{-1} \,  , \nonumber \\
&{\tilde G}^{-1}= {\tilde \gamma}- \gamma^T {\bar \gamma}^{-1} \gamma \, ,
&{\bar G}^{-1}= {\bar \gamma}- \gamma {\tilde \gamma}^{-1} \gamma^T \,  .
\end{eqnarray}

We will also use the expression similar to  the effective metric (\ref{eq:gdef}) and non-commutativity  parameters
 but with all contributions from $ab$ subspace
\begin{equation}\label{eq:ghat}
{\hat g}_{ab} = ({\tilde G} -4 {\tilde b}  {\tilde G}^{-1} {\tilde b})_{ab}   \, , \qquad
 {\hat \theta}_0^{ab} = -\frac{2}{\kappa} ({\hat g}^{-1} {\tilde b} {\tilde G}^{-1})^{ab}   \, .
\end{equation}
Note that ${\hat g}_{ab} \neq {\tilde g}_{ab}$ because ${\tilde g}_{ab}$ is projection of $g_{\mu \nu}$ on subspace $ab$.
It is extremely useful to introduce background field combinations
\begin{equation}\label{eq:bfc}
\Pi_{0 \pm ab}= b_{ab} \pm  \frac{1}{2} G_{ab}   \,  \qquad
{\hat \theta}^{ab}_{0 \pm} =  -\frac{2}{\kappa} ({\hat g}^{-1} {\tilde \Pi}_{0 \pm} {\tilde G}^{-1})^{ab} =
{\hat \theta}_0^{ab} \mp \frac{1}{\kappa}  ({\hat g}^{-1})^{ab} \, ,
\end{equation}
which are inverse to each other
\begin{equation}\label{eq:inv}
{\hat \theta}^{ac}_{0 \pm}  \Pi_{0 \mp cb} = \frac{1}{2 \kappa} \delta^a_b   \, .
\end{equation}
The similar relations valid in $ij$ subspace.

With the help of  (\ref{eq:orwc})  one can prove the relation
\begin{eqnarray}\label{eq:usr}
({\tilde g}^{-1} \beta_1  D^{-1})^a {}_i= ({\hat g}^{-1} \beta_1 {\bar \gamma}^{-1})^a {}_i  \, ,
\end{eqnarray}
where $D^{ij}$ is defined in (\ref{eq:AD1}).


\subsection{Block-wise expressions for weakly curved background fields}

For the effective metric tensor and the Kalb-Ramond background fields (\ref{eq:risp}) we define
\begin{equation}\label{eq:dgm}
{}^{\diamond} G^E_{\mu \nu} = \left (
\begin{array}{cc}
{}^{\diamond} {\tilde G}^E_{ab}    &   {}^{\diamond} G^E_{aj}       \\
 {}^{\diamond} G^E_{ib}            &  {}^{\diamond} {\bar G}^E_{ij}
\end{array}\right )   \equiv
 \left (
\begin{array}{cc}
{}^{\diamond} {\tilde G}_E    &   {}^{\diamond} G^T_E       \\
{}^{\diamond} G_E            &   {}^{\diamond} {\bar G}_E
\end{array}\right )   \, ,
\end{equation}
and
\begin{equation}
{}^{\diamond} B_{\mu \nu} = \left (
\begin{array}{cc}
{}^{\diamond} {\tilde B}_{ab}    &   {}^{\diamond} B_{aj}       \\
{}^{\diamond} B_{ib}            &  {}^{\diamond} {\bar B}_{ij}
\end{array}\right )   \equiv
 \left (
\begin{array}{cc}
{}^{\diamond} {\tilde B}    &    - {}^{\diamond} B^T       \\
 {}^{\diamond} B            &  {}^{\diamond} {\bar B}
\end{array}\right )   \, .
\end{equation}

It is also useful to introduce new notation for expressions
\begin{equation}\label{eq:dbgnmj}
({}^{\diamond} B G^{-1})_\mu{}^\nu  = \left (
\begin{array}{cc}
{}^{\diamond} {\tilde B}  {\tilde \gamma}- {}^{\diamond} B^T \gamma      &   {}^{\diamond} {\tilde B}\gamma^T- {}^{\diamond} B^T {\bar \gamma}         \\
 {}^{\diamond} B  {\tilde \gamma} + {}^{\diamond} {\bar B} \gamma       &    {}^{\diamond} B \gamma^T + {}^{\diamond} {\bar B} {\bar \gamma}
\end{array}\right )   \equiv
 \left (
\begin{array}{cc}
{}^{\diamond} {\tilde \beta}    &  {}^{\diamond}  \beta_1       \\
{}^{\diamond} \beta_2            &  {}^{\diamond} {\bar \beta}
\end{array}\right )   \, ,
\end{equation}
and
\begin{equation}\label{eq:gnmjdb}
(G^{-1} \, {}^{\diamond} B)^\mu{}_\nu  = \left (
\begin{array}{cc}
 {\tilde \gamma} {}^{\diamond} {\tilde B} + \gamma^T  {}^{\diamond} B      &   - {\tilde \gamma} \, {}^{\diamond} B^T + \gamma^T {}^{\diamond} {\bar B}         \\
 \gamma  \, {}^{\diamond} {\tilde B} + {\bar \gamma} {}^{\diamond} B       &    -  \gamma  \, {}^{\diamond} B^T + {\bar \gamma} {}^{\diamond} {\bar B}
\end{array}\right )   \equiv
 \left (
\begin{array}{cc}
- {}^{\diamond} {\tilde \beta}^T    &    - {}^{\diamond} \beta_2^T       \\
 - {}^{\diamond} \beta_1^T            &  - {}^{\diamond} {\bar \beta}^T
\end{array}\right )   \, .
\end{equation}

We will use the effective metric and non commutativity parameter but with all contributions from $ab$ subspace
\begin{equation}\label{eq:ghatwc}
{}^{\diamond} {\hat G}^E_{ab} = ({\tilde G} -4 {}^{\diamond} {\tilde B}  {\tilde G}^{-1} {}^{\diamond} {\tilde B})_{ab}   \, ,  \qquad
{}^{\diamond} {\hat \theta}^{ab}= -\frac{2}{\kappa} [({}^{\diamond} {\hat G}_E)^{-1} {}^{\diamond} {\tilde B} {\tilde G}^{-1}]^{ab}    \, .
\end{equation}
Note that ${}^{\diamond} {\hat G}^E_{ab} \neq {}^{\diamond} {\tilde G}^E_{ab}$ because ${}^{\diamond} {\tilde G}^E_{ab}$ is projection of ${}^{\diamond}  G^E_{\mu \nu}$ on subspace $ab$.
It is also useful to introduce background field combinations in weakly curved background
\begin{equation}\label{eq:bfcwc}
{}^{\diamond} \Pi_{\pm ab}= {}^{\diamond} B_{ab} \pm  \frac{1}{2} {}^{\diamond} G_{ab}   \,  \qquad
{}^{\diamond} {\hat \theta}^{ab}_\pm =  -\frac{2}{\kappa} ( {}^{\diamond} {\hat G_E}^{-1} {}^{\diamond} {\tilde \Pi}_\pm {\tilde G}^{-1})^{ab} =
{}^{\diamond} {\hat \theta}^{ab} \mp \frac{1}{\kappa}  ({}^{\diamond} {\hat G_E}^{-1})^{ab} \, ,
\end{equation}
which are inverse to each other
\begin{equation}\label{eq:invwc}
{}^{\diamond} {\hat \theta}^{ac}_\pm  \,\, {}^{\diamond} \Pi_{\mp cb} = \frac{1}{2 \kappa} \delta^a_b   \, .
\end{equation}
The similar relations valid in $ij$ subspace.


\end{document}